\documentclass[floats,floatfix,amssymb,prd,twocolumn,superscriptaddress,nofootinbib]{revtex4-1}

\usepackage{graphicx,amssymb,amsmath,amsthm,amsfonts,epsfig,epsf}
\usepackage[linktocpage]{hyperref}
\usepackage[usenames]{color}
\usepackage{epstopdf}
\usepackage{bm}
\usepackage{dcolumn}
\usepackage[latin1]{inputenc}
\usepackage{latexsym}
\usepackage{rotating}
\usepackage{hyperref}
\usepackage{color}
\usepackage{longtable}

\usepackage{enumerate}
\usepackage{tensor,multirow}
\usepackage{url}
\usepackage{tikz,tikz-3dplot}

\usepackage{slashed}

\newcommand{\GSSI}{Gran Sasso Science Institute (GSSI), I-67100 L'Aquila, Italy}
\newcommand{\GranSasso}{INFN, Laboratori Nazionali del Gran Sasso, I-67100 Assergi, Italy}

\begin{document}
\title{Inspiralling compact objects with generic deformations}

\author{Nicholas Loutrel}
\affiliation{Dipartimento di Fisica, ``Sapienza'' Universit\`a di Roma \& Sezione INFN Roma1, Piazzale Aldo Moro 5, 00185, Roma, Italy}

\author{Richard Brito}
\affiliation{CENTRA, Departamento de F\'{\i}sica, Instituto Superior T\'ecnico -- IST, Universidade de Lisboa -- UL, Avenida Rovisco Pais 1, 1049 Lisboa, Portugal}
\affiliation{Dipartimento di Fisica, ``Sapienza'' Universit\`a di Roma \& Sezione INFN Roma1, Piazzale Aldo Moro 5, 00185, Roma, Italy}

\author{Andrea Maselli}
\affiliation{\GSSI}
\affiliation{\GranSasso}

\author{Paolo Pani}
\affiliation{Dipartimento di Fisica, ``Sapienza'' Universit\`a di Roma \& Sezione INFN Roma1, Piazzale Aldo Moro 5, 00185, Roma, Italy}

\begin{abstract}
Self-gravitating bodies can have an arbitrarily complex shape, which implies a much richer multipolar structure than that of a black hole in General Relativity. With this motivation, we study the corrections to the dynamics of a binary system due to generic, nonaxisymmetric mass quadrupole moments to leading post-Newtonian (PN) order. Utilizing the method of osculating orbits and a multiple scale analysis, we find analytic solutions to the precession and orbital dynamics of a (generically eccentric) binary in terms of the dimensionless modulus parameters $\epsilon_{m}$, corresponding to axial $m=1$ and polar $m=2$ corrections from oblateness/prolateness. The solutions to the precession dynamics are exact for $0 \le \epsilon_{2} < 1$, and perturbative in $\epsilon_{1} \ll 1$. We further compute the leading order corrections to the gravitational wave amplitude and phase for a quasi-circular binary due to mass quadrupole effects. Making use of the stationary phase approximation and shifted uniform asymptotics (SUA), the corrections to the phase enter at relative 2PN order, while the amplitude modulations enter at -0.5PN order with a SUA amplitude correction at 3.25PN order, relative 2PN order to the leading order SUA correction. By investigating the dephasing due to generic quadrupole moments, we find that a phase difference $\gtrsim 0.1$~radians is achievable for $\epsilon_{m} \gtrsim 10^{-3}$, which suggests that constraints with current and future ground-based gravitational wave detectors are possible. Our results can be implemented in parameter estimation studies to constrain generic multipolar deformations of the Kerr geometry and of neutron stars.
\end{abstract}

\maketitle

\section{Introduction}\label{sec:intro}

The multipolar expansion provides a powerful tool, widely used 
in classical field theories, to characterize the distribution 
of non-symmetric distributions of charges~\cite{Jackson} and 
matter~\cite{PoissonWill}.
In General Relativity (GR) two classes of multipole 
moments can 
be defined (which are order-$\ell$ tensors): the mass moments 
${\cal Q}_{\ell m}$ and the current moments ${\cal S}_{\ell m}$ 
(henceforth $|m|\leq \ell$ is the azimuthal number of the 
multipolar decomposition and we use units in which $G=1$). 
Current moments do not have a Newtonian analogue since they 
are associated with the gravitational field produced by 
velocity fields.

Vacuum, stationary black hole~(BH) solutions in GR are also 
asymmetric and uniquely described by the Kerr 
metric~\cite{Carter:1971zc,Hawking:1971vc,Robinson:1975bv}. 
The multipole moments of a Kerr BH  satisfy closed form, elegant 
relations
\begin{align}
{\cal Q}_{\ell}&=M(ia)^\ell N_\ell \quad \ell=2,4\ldots\nonumber\\
{\cal S}_{\ell}&=iM(ia)^\ell  N_\ell \quad \ell=1,3\ldots\ ,\label{nohair}
\end{align}
where $N_\ell$ is a normalization factor~\cite{RevModPhys.52.299},
${\cal Q}_\ell \equiv {\cal Q}_{\ell 0}$ 
and ${\cal S}_\ell \equiv {\cal S}_{\ell 0}$, 
with $M\equiv {\cal Q}_0$, and ${\cal S}_1\equiv a M$ being 
the BH's mass and spin.
Thus, the multipole moments of Eqns.~\eqref{nohair} are entirely 
determined in terms of the BH's mass and spin, as dictated by the 
no-hair theorems~\cite{Carter:1971zc,Hawking:1973uf} (see 
also~\cite{Heusler:1998ua,Chrusciel:2012jk,Robinson:1975bv,Geroch:1970cd,Hansen:1974zz}). 
All other moments, namely the odd (even) $\ell$-components 
for the mass (current) multipoles, as well as the $m\neq 0$ 
terms, vanish, as a consequence of axisymmetry and of 
equatorial symmetry.

On the other hand, the fact that all multipoles with 
$\ell\geq 2$ are proportional to (powers of) the spin --~as 
well as their specific spin dependence~-- is a peculiarity of 
the Kerr metric (although not necessarily unique to Kerr~\cite{Bonga:2021ouq}). Finally, when non-spinning, 
any isolated BH must be spherically 
symmetric and described by the Schwarzschild spacetime. 

The remarkable simplicity of BHs represents an exception 
though, not shared by other self-gravitating bodies in the 
Universe.
For example, since no-hair theorems do not generically apply 
in the presence of matter, there is no compelling reason 
preventing a star from being arbitrarily deformed away 
from spherical symmetry, even when non-spinning. 
The Earth itself has a complex shape, different 
from an ellipsoid~\cite{essd-11-647-2019}.

While self-gravitating perfect fluids in a static configuration 
do not support deviations from spherical 
symmetry~\cite{1994CMaPh.162..123L}, this might not be the 
case for elastic materials~\cite{Raposo:2020yjy}.
Furthermore, it was recently shown that exotic compact 
objects can break the symmetries of a Kerr BH and have 
a much richer structure~\cite{Raposo:2018xkf,Cardoso:2019rvt}. 
In particular, smoking gun evidences for the ``non-Kerrness'' 
of a compact object would be given by the presence of 
moments that break the equatorial symmetry
(e.g. the current quadrupole ${\cal S}_2$ or the mass octopole 
${\cal Q}_3$~\cite{Fransen:2022jtw}), and/or the axisymmetry (e.g. a generic mass 
quadrupole tensor ${\cal Q}_{2m}$ with three independent 
components, $m=0,1,2$), as in the case
of multipolar boson stars~\cite{Herdeiro:2020kvf} and of 
fuzzball microstate 
geometries~\cite{Bena:2020see,Bianchi:2020bxa,Bena:2020uup,Bianchi:2020miz,Bah:2021jno}.

Checking whether such symmetry properties hold for 
an astrophysical dark object provides an opportunity 
to perform multiple null-hypothesis tests of the Kerr 
metric. The independent measurement of three 
multipole moments such as the mass, spin, and (axisymmetric) mass 
quadrupole $\mathcal{Q}_{2}$, would, for example, serve as a 
genuine strong-gravity test of the uniqueness of the 
Kerr family~\cite{Ryan:1995wh,Ryan:1997hg,Psaltis:2008bb,Gair:2012nm,Yunes:2013dva,Berti:2015itd,
Cardoso:2016ryw,Barack:2018yly,Cardoso:2019rvt}. 
In this context it is intriguing that current 
gravitational-wave~(GW) observations (especially the recent GW190814~\cite{LIGOScientific:2020zkf} and 
GW190521~\cite{Abbott:2020tfl,Abbott:2020mjq}) do not exclude the 
existence of exotic compact objects other than BHs and neutron stars. Likewise, current constraints on the spin and multipolar structure of supermassive objects coming from the Event Horizon Telescope are weak~\cite{EventHorizonTelescope:2019dse}, and do not exclude deviations from the Kerr spacetime.

The multipolar structure of a compact object leaves a footprint 
within the GW signal emitted during the coalescence of a 
binary system, by modifying at different orders the 
post-Newtonian~(PN) expansion used to model the waveform 
during the inspiral (see \cite{Blanchet:2013haa} for a review). 
Until recently, PN corrections coming from the 
multipole moments had only been computed for  
axial and equatorial symmetry, i.e. focusing on 
corrections proportional to ${\cal Q}_2$, ${\cal S}_3$ and 
${\cal Q}_4$~\cite{Ryan:1995wh,Ryan:1997hg,Poisson:1997ha,Pappas:2015npa}. 
Such calculations have been recently extended to include leading order corrections 
with broken equatorial symmetry (while preserving axisymmetry), proportional to 
${\cal S}_2$ and ${\cal Q}_3$, mostly focusing on extreme mass-ratio inspirals~(EMRIs)~\cite{Fransen:2022jtw}. 

Overall, the dominant contribution of the multipolar 
structure is encoded in the (typically spin-induced) mass 
quadrupole moment, which enters the inspiral GW phase at relative $2$PN order~\cite{Poisson:1997ha}. For comparable-mass binaries, 
this correction is expected to be measured with percent 
accuracy by third-generation ground based detectors and 
by LISA~\cite{Krishnendu:2017shb,Krishnendu:2018nqa,Kastha:2018bcr,Kastha:2019brk,Krishnendu:2019ebd}.
The PN results also provide an order of magnitude estimate 
for ``kludge'' waveforms, used to model the long 
inspiral phase of an EMRI~\cite{Barack:2006pq}. In this case 
it has been shown that LISA may constrain the mass quadrupole 
moment of the massive central object with an accuracy of one 
part in $10^4$~\cite{Barack:2006pq,Babak:2017tow}.

The aim of this paper is to extend current PN computations 
to binary configurations in which the compact objects 
show generic deformations, with no prior assumption 
on their underling symmetry. We focus in particular 
on the leading-order corrections of the mass 
quadrupole tensor ${\cal Q}_{2m}$, which enter the equations of motion to leading order at relative 2PN order. When moving into an effective one-body frame, the perturbation due to mass quadrupole effects only depends on an effective mass quadrupole moment, constituting a degeneracy between the individual moments of the compact objects.

We solve for the dynamics of the binary at relative Newtonian order, specifically we consider a reduced problem where the binary is simply described by Newtonian (or the leading PN order) dynamics, and is perturbed by the 2PN order mass quadrupole effects. We use the method of osculating orbits and multiple scale analysis to solve for the leading order corrections to the dynamics of the binary. In general, the secular dynamics of the perturbation induce precession of the orbital angular momentum. Indeed, we find that alignment between the orbital angular momentum and the Z-axis of the body can only be achieved when ${\cal{Q}}_{2,\pm1}=0$. Defining the modulus $\epsilon_{m}$ and phase $\alpha_{m}$ parameters as in Eqs.~\eqref{eq:p-params} and~\eqref{eq:a-params}, we find that the secular precession equations can be solved exactly for $\epsilon_{1} = 0$ and $0 \le \epsilon_{2} < 1$. Such solutions can be extended by working perturbatively in $\epsilon_{1} \ll 1$.

We extend the solutions to the conservative dynamics of the binary to include radiation reaction effects through the balance laws, accounting for all of the corrections due to mass quadrupole effects. Restricting to the limit of quasi-circular binaries, we compute the corrections to the TaylorF2 waveform phase using the stationary phase approximation (SPA)~\cite{Bender}. Further, we include the corrections due to orbital precession using shifted uniform asymptotics (SUA), which was originally developed for spin precessing binaries in~\cite{Klein:2014bua}. The corrections to the SPA Fourier phase enter at relative 2PN order. The amplitude modulations are controlled by the precession phase $\psi_{2}$, which enters at absolute -0.5PN order. Meanwhile, the corrections to the SUA amplitude enters at relative 3.25PN order, which is 2PN order beyond the Newtonian order SUA corrections. A simplistic estimate of the dephasing of the waveform phase suggests that small modulus values of $\epsilon_{m} \sim 10^{-3}$ might be detectable with current generation interferometers, although a detailed parameter estimation study is left for future work.

The remainder of the paper is organized as follows. In Sec.~\ref{sec:setup}, we provide an overview of the formalism we use, and some basic details of the mathematical methods needed to solve the equations of motion. In Sec.~\ref{sec:q-sol}, we solve for the conservative dynamics of the binary, specifically the secular precession effects and the leading order orbital corrections. The solutions are broken down into the oblate/prolate (often referred to as ``spheroidal'' for short) case with $\epsilon_{m} = 0$ in Sec.~\ref{sec:obl}, the polar case with $\epsilon_{1} = 0$ and $0 \le \epsilon_{2} < 1$ in Sec.~\ref{sec:polar}, and the axial case with $\epsilon_{2} = 0$ and $\epsilon_{1} \ll 1$ in Sec.~\ref{sec:axial}. We provide the general extension of the exact polar solution to include small $\epsilon_{1}$ in Sec.~\ref{sec:generic}. In Sec.~\ref{sec:rr}, we obtain the leading order corrections to radiation reaction effects. Finally, in Sec.~\ref{sec:sua}, we compute the SUA TaylorF2 waveform for quasi-circular binaries with generic mass quadrupole effects, with the main results being the GW Fourier phase given in Eq.~\eqref{eq:wave_phase}, and the GW amplitudes given in Eq.~\eqref{eq:wave_amps}. Throughout this work, we use units where $G=1$.

\section{Formalism}\label{sec:setup}

\subsection{Notation and conventions}\label{sec:notation}
We follow the same notation as in Ref.~\cite{Abdelsalhin:2018reg}, briefly summarized here. We denote the speed of light in vacuum by $c$ throughout the paper. Latin indices $i,j,k$, etc.\ run over three-dimensional spatial coordinates and are contracted with the Euclidean flat
metric $\delta^{ij}$. Since there is no distinction between upper and lower spatial indices, we will use only the upper ones throughout the paper. The
totally antisymmetric Levi-Civita symbol is denoted by $\epsilon^{ijk}$. Following the STF notation~\cite{Thorne:1980ru},
we use capital letters in the middle of the alphabet $L,K$, etc.\ as shorthand for multi-indices $a_1 \dots a_l$, $b_1
\dots b_k$, etc. Round $(\ )$, square $[\ ]$, and angular $\langle \ \rangle$ brackets in the indices indicate
symmetrization, antisymmetrization and trace-free symmetrization, respectively. 
For instance,
\begin{equation}
T^{\langle ab \rangle} = T^{(ab)} - \frac{1}{3} \delta^{ab} T^{cc} = \frac{1}{2} \left( T^{ab} + T^{ba} \right) -
\frac{1}{3} \delta^{ab} T^{cc}\,.
\end{equation}
We call \emph{symmetric trace-free} (STF) those tensors $T^{i_1\dots i_l}$ that are symmetric on all indices and
whose contraction of any two indices vanishes
\begin{align}
T^{(i_1\dots i_l)}  &= T^{i_1\dots i_l}\,,\nonumber \\
T^{i_1\dots i_k i_k \dots i_l} &= 0\,, \nonumber\\
T^{\langle i_1\dots i_l \rangle }  &= T^{i_1\dots i_l} \,. 
\end{align}
The contraction of a STF tensor $T^L$ with a generic tensor $U^L$ is $T^L U^L= T^L U^{\langle L \rangle}$. For a generic
vector $u^i$ we define $u^{ij\dots k} \equiv u^i u^j \dots u^k$ and $u^2 \equiv u^{ii}$. Derivatives with respect to the
coordinate time $t$ are expressed by overdots.

For a generic body $A$, the mass and current STF multipole tensors are denoted by $Q^L_A$ and $S^L_A$, respectively.
Restricted to a two-body system, $A =1,2$, we define the mass ratios $\eta_A = M_A/M$, where $M = M_1 + M_2 $ is the total mass and $M_A$ is the mass monopole; $M_A = Q_A$ in the Newtonian limit. The symmetric mass ratio is
$\nu = \eta_1 \eta_2$ and the reduced mass is $\mu = \nu M$. We define the dimensionless spin parameters $\chi_A = c
S_A/(\eta_A M)^2$, where $S_A = \sqrt{S^i_A S^i_A}$ is the absolute value of the current dipole moment.
The body
position, velocity and acceleration vectors are denoted by $z_A^i$, $v_A^i = \dot{z}_A^i$ and $a_A^i = \ddot{z}_A^i$,
respectively. We define the two-body relative position, velocity and acceleration vectors by $z^i = z_2^i - z_1^i$, $v^i
= v_2^i - v_1^i$ and $a^i = a_2^i - a_1^i$, respectively. We also define the relative unit vector $n^i = z^i/r$, where
$r = \sqrt{z^i z^i}$ is the orbital distance. Using these definitions the radial velocity is given by $\dot r=v^in^i$. 
%
Finally, for a binary system in circular orbit we define the PN expansion parameter
$\tilde{u} = (2\pi F M)^{1/3}/c$, where $F$ is the orbital frequency. Note that $\tilde{u} = v + {\cal{O}}(c^{-4})$.

\subsection{Main equations}\label{subsec:summrel}

The post-Newtonian Lagrangian describing the two-body interaction, up to the relevant multipole moments, can be written as
\begin{equation}
\label{eq:totallagrangian}
{\mathcal{L}} = {\cal{L}}_{\rm pp} + {\cal{L}}_{\rm spin} + {\cal{L}}_{\rm quad}\,.
\end{equation}
Here, $\mathcal{L}_{\rm pp}$ describes the PN interaction between 
two point particles of mass $m_{1}$ and $m_{2}$, which up to 
2PN order is given in relative coordinates
\begin{align}
\mathcal{L}_{\rm pp} =& {\cal{L}}_{\rm N} + c^{-2} {\cal{L}}_{\rm 1PN} + c^{-4} {\cal{L}}_{\rm 2PN} + {\cal{O}}\left(c^{-6}\right)
\,,\label{eq:LM}
\end{align}
with
\begin{align}
    {\cal{L}}_{\rm N} &= \frac{1}{2} \mu v^{2} + \frac{\mu M}{r}
\end{align}
and the higher PN order terms given in Appendix~\ref{pn}. 
The term ${\cal{L}}_{\rm spin}$ contains the contributions from current dipoles, specifically the spin angular momenta of each body, and is given to second order in spins by~\cite{PhysRevD.47.R4183}
\begin{align}
{\cal{L}}_{\rm spin} =& \frac{1}{2} \eta_{1} \eta_{2} \epsilon^{ijk} v^{i} a^{j} \Sigma^{k}  + \frac{2M}{r^{2}} \eta_{1} \eta_{2} \epsilon^{ijk} v^{i} n^{j} (S^{k}+\Sigma^{k})
\nonumber \\
&-\frac{3}{r^{3}} S^{i}_{1} S^{j}_{2} n^{<ij>}
  \ ,\label{eq:LJ}
\end{align}
where $S^{i} = S_{1}^{i} + S_{2}^{i}$ and $\Sigma^{i} = (\eta_{2}/\eta_{1}) S_{1}^{i} + (\eta_{1}/\eta_{2}) S_{2}^{i}$, with $S_{1,2}^{i}$ the spin angular momenta of each body. Finally, the mass quadrupole contribution reads, to leading PN order~\cite{Abdelsalhin:2018reg},
\begin{align}
  \mathcal{L}_{\rm quad} =& \frac{3 M}{2r^3} Q^{ij}_{\rm eff} n^{<ij>} +O \left( c^{-2} \right)  \,,
  \label{eq:LMQ}
\end{align}
where $Q^{ij}_{\rm eff} = \eta_{2} Q^{ij}_{1} - \eta_{1} Q^{ij}_{2}$, with $Q^{ij}_{1,2}$ the mass quadrupole moments of each body. In the following we shall ignore the tidal deformability of the bodies, which corresponds to the part of the Lagrangian describing the internal dynamics.

In the case of a binary system, the dynamics in the center-of-mass (COM) frame is described by the orbital separation $z^i=z^i_2-z^i_1$. From the variation of the above Lagrangian with respect to the worldline coordinates $z_{1,2}^{i}$ we can derive the equations of motion of the binary:
\begin{align}
  a^i&={\ddot z}^i={\ddot z}^i_2-{\ddot z}^i_1\nonumber\\
  &=a^i_{\rm pp}+a^i_{\rm spin}+a^i_{\rm quad}\label{eq:orbitaleom}\,.
\end{align}
The mass and spin contributions are
\begin{align}
    \label{eq:ppeom}
  a^i_{\rm pp}=&
  -\frac{M}{r^2} n^i + a^{i}_{\rm 1PN} + a^{i}_{\rm 2PN} + {\cal{O}}(c^{-5/2})\,,
  \\
  \label{eq:spineom}
  a^i_{\rm spin}=&
  \frac{1}{r^{3}} \left\{6 n^{i} \left[\epsilon^{jkp} n^{k} v^{p} (S^{j} + \Sigma^{j})\right] 
  \right.
  \nonumber\\
  &\left.
  - \epsilon^{ijk} \left[v^{j}(4S^{k} + 3\Sigma^{k}) + 3 \dot{r} n^{j}(2S^{k}+\Sigma^{k})\right]\right\}
  \nonumber \\
  &+\frac{15}{\mu r^{4}} S_{1}^{j} S_{2}^{k} n^{<ijk>} + {\cal{O}}(c^{-5/2})\,,
\end{align}
whereas the mass quadrupole contribution is
\begin{equation}
\label{eq:quadmass}
    a^i_{\rm quad}=
     -\frac{15 Q^{<jk>}_{\rm eff}}{2 \nu r^4} n^{<ijk>} + O(c^{-6})    \,.
\end{equation}

The orbital equations of motion must be supplemented by a suitable set of equations describing the dynamics of the spin angular momenta of each body. There are two ways of achieving this, through the fluid description of PN sources~\cite{PoissonWill} or through effective field theory~\cite{Steinhoff:2011sya}. Both methods give the same result, namely
\begin{equation}
\label{eq:dSdt}
    \frac{dS^{i}_{1}}{dt} = \epsilon^{ijk} \left[\Omega^{j} S_{1}^{k} + {\cal{T}}^{jk}_{\rm QM}\right]\ ,
\end{equation}
where $S^{i}_{1}$ is the spin of one of the bodies, $\Omega^{j}$ contains the spin-orbit and spin-spin couplings
\begin{equation}
    \Omega^{j}=c^{-3/2} \Omega_{\rm SO}^{j} + c^{-4} \Omega_{\rm SS}^{j} + {\cal{O}}(c^{-5/2})\,,
\end{equation}
and ${\cal{T}}_{\rm QM}^{jk}$ is the torque generated by the monopole-quadrupole interaction
\begin{equation}
    {\cal{T}}^{jk}_{\rm QM} = \frac{3\eta_{2} M}{r^{3}} Q_{1}^{<ja>} n^{<ka>}\,.
\end{equation}
The spin precession equation for the other body can be found by taking $(1\leftrightarrow2)$ in Eq.~\eqref{eq:dSdt}.

The coupled system of orbital equations of motion and spin precession equations possess constants of motion. The first is associated with the fact that the Lagrangian in Eq.~\eqref{eq:totallagrangian} is explicitly time independent, and thus has a conserved Hamiltonian ${\cal{H}}= p^{i}v^{i} + s^{i}a^{i} - {\cal{L}}$, where $p^{i} = \partial{\cal{L}}/\partial v^{i}$ and $s^{i} = \partial{\cal{L}}/\partial a^{i}$. This leads to the conserved orbital energy of the binary
\begin{align}
    \label{eq:Eorb}
    E_{\rm orb} &= E_{\rm N} + E_{\rm quad} + c^{-2} E_{\rm 1PN} + c^{-3/2} E_{\rm SO} 
    \nonumber\\
    &+ c^{-4} \left[E_{\rm 2PN} + E_{\rm SS} \right] + {\cal{O}}(c^{-5/2})\ ,
\end{align}
where
\begin{align}
    E_{\rm N} &= \frac{1}{2} \mu v^{2} - \frac{\mu M}{r}\ ,
    \\
    E_{\rm quad} &= - \frac{3M}{2r^{3}} Q_{\rm eff}^{ij} n^{<ij>}\ ,
\end{align}
and the remaining PN terms are given in Appendix~\ref{pn}. In addition, one can define the conserved total angular momentum $J^{i} = \epsilon^{ijk} (r^{i} p^{k} + v^{i} s^{k}) + S^{i}$. The first of these terms constitutes the orbital angular momentum, which is
\begin{align}
\label{eq:Lorb}
    L^{i} &= \mu \epsilon^{ijk} r^{j} v^{k} \left[1 + c^{-2} L_{\rm 1PN} + {c^{-4}} L_{\rm 2PN}\right] 
    \nonumber \\
    &+ c^{-3/2} L^{i}_{\rm SO} 
     + {\cal{O}}(c^{-5/2})\,.
\end{align}
The first term in the above is the Newtonian orbital angular momentum, while the remaining PN and spin-orbit terms are given in Appendix~\ref{pn}. Note that the direction $\hat{L}^{i} = L^{i}/L$ is not conserved and obeys the precession equation
\begin{equation}
\label{eq:dLdt}
    \frac{d\hat{L}^{i}}{dt} = L^{-1} \frac{dS^{i}}{dt}\,,
\end{equation}
where $L=\sqrt{L^i L^i}$.
The set of equations has now been completed.

Before continuing, is it worth pointing out an additional property of the orbital and spin angular momenta in the case of generic quadrupole effects. In the absence of radiation reaction, the magnitude of these angular momenta, specifically $L$ given in Eq.~\eqref{eq:Lorb} and $S_{A} = \sqrt{S_{A}^{i} S_{A}^{i}}$, are not conserved when considering generic quadrupole corrections. This may seem rather confusing when comparing to the well studied scenario of spin precessing BHs, where the quadrupole moment scales as $Q_A^{ij} \propto S^{i}_{A} S^{j}_{A}$. It is well known from the PN spin precession equations that $S_{A}$ is conserved in this case~\cite{Steinhoff:2011sya}. However, a quick contraction of Eq.~\eqref{eq:dSdt} with $S_{1}^{i}$ reveals that the spin magnitude is only conserved when $\epsilon^{ijk} Q^{<ja>} n^{<ka>} \perp S^{i}$, which need not be true for an arbitrary quadrupole moment. A similar result can be found for the conservation of $L$. While this may seem problematic, it is important to remember that the conserved quantity is actually $J = \sqrt{J^{i}J^{i}}$, which will change only when we include radiation reaction.

\subsection{Osculating orbits}
\label{sec:osculat}

\begin{figure}[htb!]
    \includegraphics[trim=6cm 4cm 6cm 3cm, clip, width=\columnwidth]{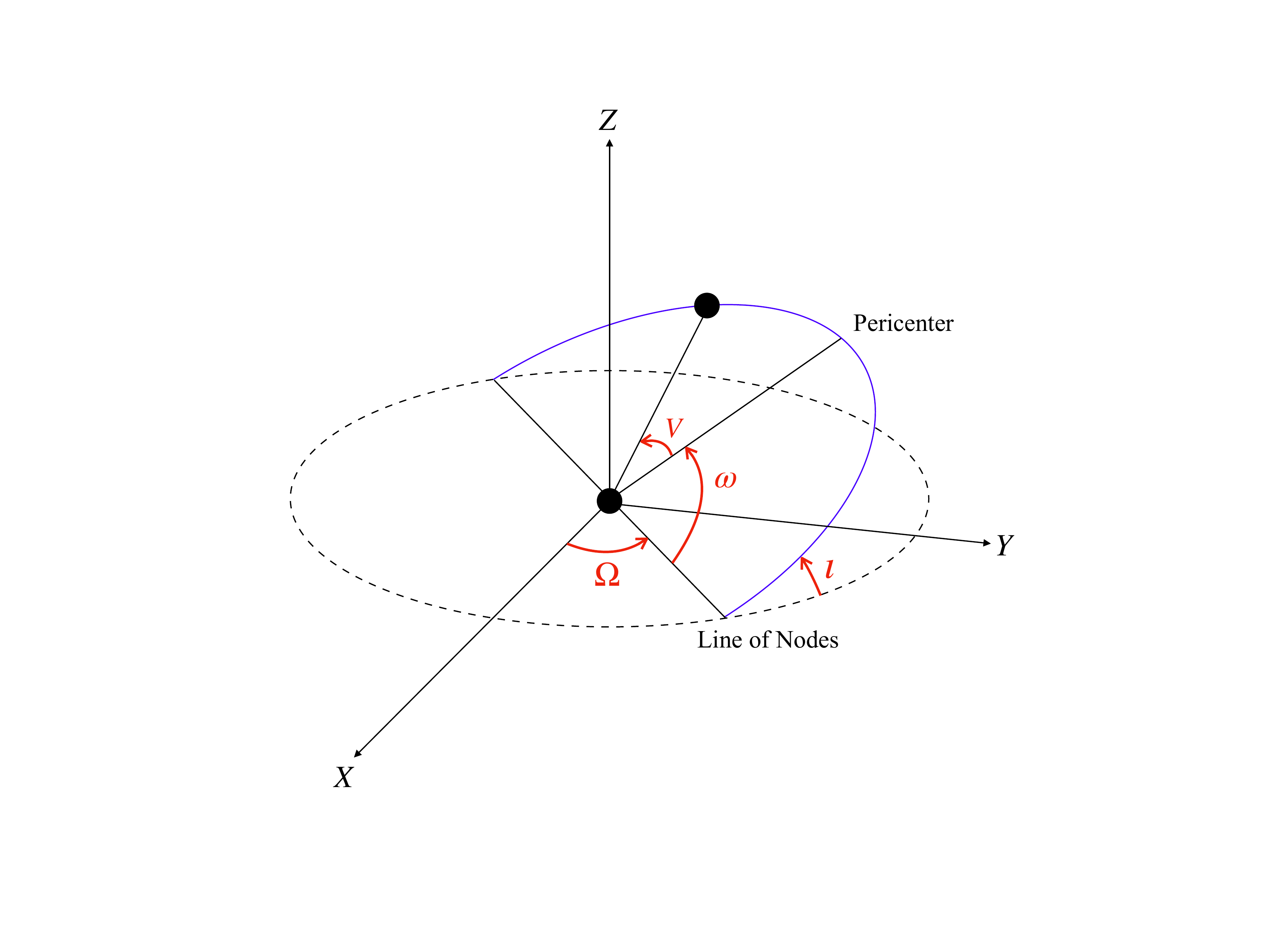}
    \caption{Graphical sketch of the orbital motion (blue solid line) 
as viewed in the fundamental reference frame. Here $\iota$ is the 
inclination angle, $\Omega$ is the longitude of the ascending 
node, $\omega$ is the longitude of pericenter and $V$ is the 
true anomaly.}\label{fig:orbit}
\end{figure}

Consider the problem of solving for the orbital motion which is described by Eq.~\eqref{eq:orbitaleom}. To simplify the calculation, consider the case of Newtonian orbits acted upon by a perturbing force generated from the higher PN terms, spin, and mass quadrupole effects, given by the 1PN and 2PN terms in Eq.~\eqref{eq:ppeom} and Eqs.~\eqref{eq:spineom}-\eqref{eq:quadmass}, respectively. In such a case, the action of the perturbing force can be understood using the method of osculating orbits~\cite{PoissonWill}. For unperturbed Newtonian orbits, the equations governing any bound orbit are
\begin{align}
\vec{r} &= r_{K} \hat{n}\,,
\\
\vec{v} &= \dot{r}_{K} \hat{n} + r \dot{\phi}_{K} \hat{\lambda}\,,
\end{align}
where we recall that $r$ is the relative radial distance, $v$ is the relative velocity, $\phi$ is the orbital phase, and $[\hat{n},\hat{\lambda}]$ are the basis vectors that parameterize the orbital plane. The Keplerian expressions for the radial separation and velocities are
\allowdisplaybreaks[4]
\begin{align}
    \label{eq:r-kep}
    r_{K} &= \frac{p}{1+e \cos V}\,,
    \\
    \dot{r}_{K} &= e \left(\frac{M}{p}\right)^{1/2} \sin V\,,
    \\
    \label{eq:phidot-kep}
    \dot{\phi}_{K} &= \left(\frac{M}{p^{3}}\right)^{1/2} \left(1 + e \cos V\right)^{2}\,,
\end{align}
where $e$ is the Newtonian orbital eccentricity, $p$ is the semi-latus rectum, and $V = \phi - \omega$ is the true anomaly, with $\omega$ the longitude of pericenter. In the absence of a perturbing force, the motion of the binary is planar. However, this is not necessarily true in the perturbed case, and we need to generalize the prescription of the orbit further. Defining a new frame spanned by the vectors $[\hat{e}_{X}, \hat{e}_{Y}, \hat{e}_{Z}]$, the orbit can be arranged into an arbitrary orientation with respect to this new frame through the use of Euler angles, as illustrated in Fig.~\ref{fig:orbit}. A sufficient parameterization is~\cite{PoissonWill}
\allowdisplaybreaks[4]
\begin{widetext}
\begin{align}
    \label{eq:n-vec}
    \hat{n} &= [\cos\Omega \cos\phi - \cos \iota \sin\Omega \sin\phi, \sin\Omega \cos\phi + \cos\iota \cos\Omega \sin\phi, \sin\iota \sin\phi]\,,
    \\
    \hat{\lambda} &= [-\cos\Omega \sin\phi - \cos\iota \sin\Omega \cos\phi, -\sin\Omega \sin\phi + \cos\iota \cos\Omega \cos\phi, \sin\iota \cos\phi]\,,
    \\
    \label{eq:L-vec}
    \hat{L} &= [\sin\iota \sin\Omega, -\sin\iota \cos\Omega, \cos\iota]\,,
\end{align}
\end{widetext}
where $\iota$ is the inclination angle and $\Omega$ is the longitude of the ascending node. The Keplerian orbit is now parameterized by five conserved quantities $\mu^{a} = [p, e, \iota, \omega, \Omega]$.

The method of osculating orbits posits that, under the action of any perturbing force, the parameters $\mu^{a}$ are no longer constant, but vary in time according to the perturbing force. The trajectory of the binary is parameterized by $\vec{r} = \vec{r}(t, \mu^{a})$ and $\vec{v} = \vec{v}(t, \mu^{a})$, while the equations of motion are
\begin{align}
    \frac{d}{dt} \vec{r}(t, \mu^{a}) &= \vec{v}(t, \mu^{a})\,,
    \\
    \frac{d}{dt} \vec{v}(t, \mu^{a}) &= \vec{f}_{N} + \vec{f}_{\rm pert}\,,
\end{align}
with $\vec{f}_{\rm pert}$ the perturbing force. The method of osculating orbits promotes the conserved parameters to functions of the time variable, specifically $\mu^{a} \rightarrow \mu^{a}(t)$, and thus
\begin{equation}
    \frac{d}{dt} = \frac{\partial}{\partial t} + \frac{d\mu^{a}}{dt} \frac{\partial}{\partial \mu^{a}}\,.
\end{equation}
The first term above generates the usual conserved Keplerian orbits, while the remaining equations are
%
\begin{align}
    \label{eq:osc1}
    \frac{d\mu^{a}}{dt} \frac{\partial \vec{r}}{\partial \mu^{a}} &= 0\,,
    \\
    \label{eq:osc2}
    \frac{d\mu^{a}}{dt} \frac{\partial \vec{v}}{\partial \mu^{a}}  &= \vec{f}_{\rm pert}(\mu^{a})\,.
\end{align}
Specifying the perturbing force as $\vec{f}_{\rm pert} = {\cal{R}} \hat{n} + {\cal{S}} \hat{\lambda} + {\cal{W}} \hat{L}$, the osculating equations for $\mu^{a}$ are~\cite{PoissonWill}
\begin{widetext}
\begin{align}
\label{eq:dp-osc}
    \frac{dp}{dt} &= 2\left(\frac{p^{3}}{M}\right)^{1/2} \frac{{\cal{S}}}{1 + e \cos V}\,,
    \\
    \label{eq:de-osc}
    \frac{de}{dt} &= \left(\frac{p}{M}\right)^{1/2} \left[\sin V {\cal{R}} + \frac{2 \cos V + e (1 + \cos^{2}V)}{1 + e \cos V} {\cal{S}}\right]\,,
    \\
    \label{eq:di-osc}
    \frac{d\iota}{dt} &= \left(\frac{p}{M}\right)^{1/2} \frac{\cos(V+\omega)}{1 + e \cos V} {\cal{W}}\,,
    \\
    \label{eq:dO-osc}
    \frac{d\Omega}{dt} &= \left(\frac{p}{M}\right)^{1/2} \frac{\sin(V + \omega)}{1 + e \cos V} \frac{{\cal{W}}}{\sin\iota}\,,
    \\
    \label{eq:dw-osc}
    \frac{d\omega}{dt} &= \frac{1}{e} \left(\frac{p}{M}\right)^{1/2} \left[-\cos V {\cal{R}} + \frac{2 + e \cos V}{1 + e \cos V} \sin V {\cal{S}} - e \cot\iota \frac{\sin(V + \omega)}{1 + e \cos V} {\cal{W}}\right]\,.
\end{align}
\end{widetext}
Note that there are only five parameters $\mu^{a}$, but Eqs.~\eqref{eq:osc1}-\eqref{eq:osc2} are six equations in total. The five equations for $\mu^{a}$ are supplemented by an additional equation for the true anomaly $V$ in order to complete this system of equations. Such equation is given by $\dot{V} = \dot{\phi}_{K} - (\dot{\omega} + \dot{\Omega} \cos\iota)$, which uses the above equations for $\omega$ and $\Omega$. The action of the perturbing force on the orbit is now fully specified. 

When studying the evolution of the osculating equations, it is important to realize that they depend on at least two timescales, the orbital timescale encoded through the dependence on $V$ and a secular timescale, which is determined by the perturbing force. In order to obtain PN accurate solutions, we must then solve the osculating equations using multiple scale analysis~\cite{PoissonWill} (in this case, two timescale analysis). Because the equations are parameterized in terms of $V$ rather than $t$, it is convenient to recast them as $d\mu^{a}/dV = (d\mu^{a}/dt)/(dV/dt)$ and PN expand to the relevant order. The two scales of the problem then become $V$ which is the shorter scale, and $\tilde{V} := \epsilon V$ the longer scale, where $f^{i}_{\rm pert} = {\cal{O}}(\epsilon)$ with $\epsilon$ an order keeping parameter. The derivative operator then becomes,
\begin{equation}
    \label{eq:msa-d}
    \frac{d}{dV} = \frac{\partial}{\partial V} + \epsilon \frac{\partial}{\partial \tilde{V}}\,,
\end{equation}
and our ansatz for the solution is
\begin{equation}
    \label{eq:msa-sol}
    \mu^{a} = \mu^{a}_{0}(\tilde{V}) + \epsilon \mu^{a}_{1}(V,\tilde{V}) + {\cal{O}}(\epsilon^{2})\,.
\end{equation}
The leading order term above $\mu^{a}_{0}$ is only dependent on the long secular scale ${\tilde{V}}$, since the $\mu^{a}$ are conserved for unperturbed Keplerian orbits. 

The strategy to solve the osculating equations is to combine Eqs.~\eqref{eq:msa-d}~--~\eqref{eq:msa-sol} with Eqs.~\eqref{eq:dp-osc}~--~\eqref{eq:dw-osc}, and expand to the relevant order in $\epsilon$. The leading order equation is
\begin{equation}
    \label{eq:msa-eq}
    \frac{d\mu^{a}_{0}}{d\tilde{V}} + \frac{\partial \mu^{a}_{1}}{\partial{V}} = {\cal{F}}^{a}(V ; \mu^{a}_{0})\,,
\end{equation}
where ${\cal{F}}^{a}$ are given by the right hand side of Eqs.~\eqref{eq:dp-osc}~--~\eqref{eq:dw-osc}. This equation can be solved by realizing that the dependence on the shorter scale $V$ is purely oscillatory. Upon averaging in the following fashion,
\begin{equation}
    \langle f \rangle = \frac{1}{2\pi} \int_{0}^{2\pi} f(V) dV\,,
\end{equation}
Eq.~\eqref{eq:msa-eq} reduces to 
\begin{equation}
    \label{eq:msa-avg}
    \frac{d\mu^{a}_{0}}{d\tilde{V}} = \langle {\cal{F}}^{a}\rangle(\mu^{a}_{0})\,,
\end{equation}
which uniquely determines $\mu^{a}_{0}$. Finally, to obtain $\mu^{a}_{1}$, we combine Eq.~\eqref{eq:msa-avg} with Eq.~\eqref{eq:msa-eq} and integrate with respect to $V$, specifically
\begin{equation}
    \label{eq:msa-osc}
    \mu^{a}_{1}(V,\tilde{V}) = \mu^{a}_{1, {\rm sec}}(\tilde{V}) + \int dV \left[{\cal{F}}^{a}(V;\mu^{a}_{0}) - \langle {\cal{F}}^{a} \rangle (\mu^{a}_{0})\right]\,.
\end{equation}
This determines $\mu^{a}_{1}$ up to a purely secular term $\mu^{a}_{1,{\rm sec}}(\tilde{V})$, which is determined by next order equations in $\epsilon$. For the purposes of the present calculation, it suffices to stop the analysis here.

\section{Generic Mass Quadrupole Effects}
\label{sec:q-sol}

The perturbing force we desire to investigate is given in Eq.~\eqref{eq:quadmass}, which is dependent on the effective quadrupole tensor $Q_{\rm eff}^{ij}$. In order to calculate the necessary components of the perturbing force for the osculating equations, we need to specify the components of this STF tensor. To do so, we assume the quadrupole moment is held fixed with respect to the $(XYZ)$-frame, which we now refer to as the body frame. For convenience, this frame is also chosen such that the direction of the total angular momentum $J^{i}$ is aligned with the $Z$-directions $e_{Z}^{i}$. In this frame, the STF tensor can be readily decomposed into spherical harmonics, specifically
\begin{equation}
    Q^{<ij>}_{\rm eff} = W_{2} \sum_{m=-2}^{2} {\cal{Y}}_{2m}^{<ij>} Q_{m}\,,
\end{equation}
where $Q_{m}$ are the spherical harmonic coefficients of the mass quadrupole, ${\cal{Y}}_{lm}^{<L>}$ are defined as
\begin{equation}
    {\cal{Y}}_{lm}^{<L>} = \frac{1}{W_{l}} \int dS^{2} N^{<L>} Y_{lm}^{\dagger}(\theta,\phi)\,,
\end{equation}
with $N^{i} = [\sin\theta\cos\phi,\sin\theta\sin\phi,\cos\theta]$, $Y_{lm}(\theta,\phi)$ the spherical harmonic functions, $W_{l} = 4\pi l!/(2l+1)!!$, and the integral is performed over the 2-sphere. Note that in general, the $Q_{m}$'s are complex (except when $m=0$) while the components of $Q^{<ij>}$ are real. It is thus simpler to specify the components of $Q^{<ij>}$ in terms of the real and imaginary parts of the $Q_{m}$'s, specifically
\begin{align}
    Q_{+1} &= Q^{R}_{+1} + i Q^{I}_{+1}\,,
    \\
    Q_{+2} &= Q^{R}_{+2} + i Q^{I}_{+2}\,.
\end{align}
The decomposition for the negative $m$ terms follows from $Q_{-m} = (-1)^{m} Q_{m}^{\dagger}$, while $Q_{0}$ is real valued. With this, the osculating equations for the mass quadrupole correction become
\allowdisplaybreaks[4]
\begin{widetext}
\begin{align}
    \label{eq:di-Q}
    \Big\langle\frac{d\iota}{dt}\Big\rangle &= \left(\frac{6\pi}{5}\right)^{1/2} \frac{(1-e^{2})^{3/2}}{\nu M^{1/2} p^{7/2}} \left\{\cos\iota (-Q^{R}_{+1} \cos\Omega + Q^{I}_{+1} \sin\Omega) + \sin\iota [Q^{R}_{+2} \sin(2\Omega) + Q^{I}_{+2} \cos(2\Omega)]\right\}\,,
    \\
    \label{eq:dO-Q}
    \Big\langle\frac{d\Omega}{dt}\Big\rangle &= \left(\frac{\pi}{5}\right)^{1/2} \frac{(1-e^{2})^{3/2}}{\nu M^{1/2} p^{7/2}} \left\{\cos\iota [3 Q_{0} + \sqrt{6} Q^{R}_{+2} \cos(2\Omega) - \sqrt{6} Q^{I}_{+2} \sin(2\Omega)] + \sqrt{6} \cos(2\iota) \csc\iota (Q^{R}_{+1} \sin\Omega + Q^{I}_{+1} \cos\Omega)\right\}\,,
    \\
    \label{eq:dw-Q}
    \Big\langle\frac{d\omega}{dt}\Big\rangle &= \frac{1}{4} \left(\frac{\pi}{5}\right)^{1/2} \frac{(1-e^{2})^{3/2}}{\nu M^{1/2} p^{7/2}} \left\{-3 Q_{0} \left[3 + 5 \cos(2\iota)\right] + 2\sqrt{6} \left[3-5\cos(2\iota)\right] \cot\iota \left(Q^{R}_{+1} \sin\Omega + Q^{I}_{+1} \cos\Omega\right) 
    \right.
    \nonumber \\
    &\left.
    + \sqrt{6} \left[1 - 5 \cos(2\iota)\right] \left(Q^{R}_{+2} \cos(2\Omega) - Q^{I}_{+2} \sin(2\Omega)\right)  \right\}\,,
\end{align}
\end{widetext}
while $[e,p]$ do not change on the secular timescale to leading order and are determined by Eq.~\eqref{eq:msa-osc}.

The above osculating equations are actually equivalent to Eq.~\eqref{eq:dLdt}, describing the precession of the orbital angular momentum $L^{i}$ around the total angular momentum $J^{i}$. In order for the direction of $J^{i}$ to be conserved, one also has to consider the precession of the spins, which after orbit averaging Eq.~\eqref{eq:dSdt} become
\begin{align}
\label{eq:dS1-Q}
    \Big\langle \frac{dS^{i}_{1}}{dt} \Big\rangle &=  -\frac{3\eta_{2} M}{2p^{3}}(1-e^{2})^{3/2} \epsilon^{ijk} Q_{1}^{<ja>} \hat{L}^{<ka>}\,,
    \\
    \label{eq:dS2-Q}
    \Big\langle \frac{dS^{i}_{2}}{dt} \Big\rangle &= -\frac{3\eta_{1} M}{2p^{3}}(1-e^{2})^{3/2} \epsilon^{ijk} Q_{2}^{<ja>} \hat{L}^{<ka>}\,.
\end{align}
For the present calculations, we will neglect the spin-orbit and spin-spin effects when considering the precession induced by quadrupole effects. The reason for this is that the PN precession equations up to the relevant PN order have only recently been solved analytically in the case when $Q^{ij}$ corresponds to the spin-induced quadrupole moment~\cite{Chatziioannou:2016ezg,Chatziioannou:2017tdw,Racine:2008qv,Kesden:2014sla}. The case for generic $Q^{ij}$ has not been solved. 
For simplicity, we shall consider only nonspinning binaries and hence neglect the relativistic spin-orbit and spin-spin couplings. In such a scenario, the problem reduces down to first solving Eqs.~\eqref{eq:di-Q}-\eqref{eq:dw-Q}, and then solving the above spin precession equations for $S_{1,2}^{i}$. In the following sections we consider the generic problem of solving the osculating equations.

As a special application of the system of equations given by Eqs.~\eqref{eq:di-Q}-\eqref{eq:dw-Q} and Eqs.~\eqref{eq:dS1-Q}-\eqref{eq:dS2-Q}, we consider the case where $\hat{L}^{i}$ and $S_{1,2}^{i}$ are aligned with the Z-direction of the body frame in Sec.~\ref{align}.

\subsection{Precessing Solutions}
\label{sec:prec}

Consider the problem of solving Eqs.~\eqref{eq:di-Q}-\eqref{eq:dw-Q}. In general, there does not appear to be a closed-form analytic solution to this system for generic non-zero quadrupole coefficients $[Q_{0}, Q^{R,I}_{+1}, Q^{R,I}_{+2}]$ and $\iota \ne 0$. However, there are some special configurations which allow for closed-form solutions. The three cases are as follows:
\begin{itemize}
    \item Spheroidal: $Q_{\pm 1} = 0 = Q_{\pm 2}$ with $Q_{0}$ non-vanishing
    \item Polar: $Q_{\pm 1} = 0$ with $[Q_{0}, Q_{\pm 2}]$ non-vanishing
    \item Axial: $Q_{\pm 2} = 0$ with $[Q_{0}, Q_{\pm 1}]$ non-vanishing
\end{itemize}
Below, we detail each of these cases.

\subsubsection{Spheroidal Case}
\label{sec:obl}
The spheroidal case considers the scenario where the compact object has an oblate/prolate spheroidal shape, and thus the only non-vanishing quadrupole coefficients is the $m=0$ term. A common astrophysical scenario that would create such an effect is a quadrupole moment induced by rotation. For the calculation at hand, we leave $Q_{0}$ as a generic constant. However, in the case of spin-induced quadrupole moment, $Q_{0} = C_{Q} \chi^{2} M^{3}+{\cal O}(\chi^4)$ with $\chi$ the dimensionless spin parameter, $M$ the mass of the compact object, and a proportionality factor $C_{Q}$ which is dependent on the equation of state. In the Kerr BH case, $C_Q=-1$ and higher-order spin corrections in $Q_0$ vanish identically.

In this scenario, the secular equations simplify to
\begin{align}
    \label{eq:di-oblate}
    \Big \langle \frac{d\iota}{dt} \Big \rangle &= 0\,,
    \\
    \label{eq:dO-oblate}
    \Big \langle \frac{d\Omega}{dt} \Big \rangle &= 3 \left(\frac{\pi}{5}\right)^{1/2} \frac{(1-e^{2})^{3/2}}{\nu M^{1/2} p^{7/2}} Q_{0} \cos\iota\,,
    \\
    \label{eq:dw-oblate}
    \Big \langle \frac{d\omega}{dt} \Big \rangle &= -\frac{3}{4} \left(\frac{\pi}{5}\right)^{1/2} \frac{(1-e^{2})^{3/2}}{\nu M^{1/2} p^{7/2}} Q_{0} [3 + 5 \cos(2\iota)]\,.
\end{align}
As can be seen from Eq.~\eqref{eq:di-oblate}, the inclination angle becomes constant, and thus, there is no nutation. The only effect on the binary is the precession of the orbital plane, encoded through $[\Omega, \omega]$. Taking $\iota = \iota_{0} = {\rm const.}$, and defining
\begin{equation}
    \label{eq:psio}
    \frac{d\psi_{0}}{dt} = 3 \left(\frac{\pi}{5}\right)^{1/2} \frac{Q_{0} (1-e^{2})^{3/2}}{\nu M^{1/2} p^{7/2}} \cos\iota_{0}\,,
\end{equation}
Eqs.~\eqref{eq:dO-oblate}-\eqref{eq:dw-oblate} can be directly integrated to obtain
\begin{align}
    \label{eq:obl-sol}
    \Omega &= \psi_{0}\,, \qquad \omega = -\frac{1}{4} \sec\iota_{0} [3 + 5 \cos(2\iota_{0})] \psi_{0}\,.
\end{align}
Note that here we wrote the solutions in terms of the dependent variable $\psi_{0}$ instead of time. The reason for this is that the right hand side of Eq.~\eqref{eq:psio} is function of the orbital velocity through $[p,e]$ and will thus change on the radiation reaction timescale. We expand on this more in Sec.~\ref{sec:rr}.

\subsubsection{Polar Case}
\label{sec:polar}
The polar case is named due to the fact that the non-vanishing quadrupole coefficients $[Q_{0}, Q_{\pm 2}]$ correspond to spherical harmonics modes that are even under spatial reflection, i.e. polar modes. In this scenario, it is convenient to define the dimensionless parameters $\mathfrak{r}_{2} = Q^{R}_{+2}/Q_{0}$ and $\mathfrak{i}_{2} = Q^{I}_{+2}/Q_{0}$. Further, we define the \textit{polar modulus} $\epsilon_{2}$ and \textit{polar argument} $\alpha_{2}$ such that
\begin{equation}
    \label{eq:p-params}
    \epsilon_{2} = \left[\frac{2}{3} \left({\mathfrak{r}}_{2}^{2} + {\mathfrak{i}}_{2}^{2}\right)\right]^{1/2}\,, \qquad \alpha_{2} = \frac{1}{2}\tan^{-1}\left({\mathfrak{i}}_{2}/{\mathfrak{r}}_{2}\right)\,.
\end{equation}
Finally, we define $\psi_{2}$ such that
\begin{equation}
    \label{eq:psip}
    \frac{d\psi_{2}}{dt} = 3 \left(\frac{\pi}{5}\right)^{1/2} \frac{Q_{0}(1-e^{2})^{3/2}}{\nu M^{1/2} p^{7/2}} \sqrt{1 - \epsilon_{2}^{2}} \cos\iota\,.
\end{equation}
Unlike the spheroidal case, the inclination angle is no longer constant and the primary effect of the $m=\pm2$ modes is to induce nutation of the orbital angular momentum. As a result, the above definition for $\psi_{2}$ no longer varies on solely the radiation reaction timescale, but also on the precession timescale through $\iota$. Also, note the presence of $\epsilon_{2}$ in Eq.~\eqref{eq:psip}, as opposed to Eq.~\eqref{eq:psio} since $\epsilon_{2}=0$ in the spheroidal case.

The starting point for solving the secular equations in this case is to divide Eq.~\eqref{eq:dO-Q} by Eq.~\eqref{eq:psip}. Defining ${\cal{V}} = \Omega + \alpha_{2}$, we arrive at
\begin{align}
    \label{eq:dVdpsi}
    \frac{d{\cal{V}}}{d\psi_{2}} = \frac{1 + \epsilon_{2} \cos{(2\cal{V}})}{\sqrt{1 - \epsilon_{2}^{2}}}\,.
\end{align}
The solution to this equation depends on the value of $\epsilon_{2}$, which depends on the specific scenario under consideration. For astrophysical objects, the induction of a quadrupole moment on the body is largely expected to be a result of spin angular momentum creating an $m=0$ contribution. However, such objects are also likely not perfectly spheroidal, but may have small deviations on their surface (e.g. mountains on a neutron star) which could contribute to $|m|>0$ modes~\cite{Raposo:2020yjy}. Such contributions are expected to be small, and we could thus assume that $\epsilon_{2} \ll 1$. 

Another example of how to generate an $|m|>0$ mode on a compact object is through dynamical tides. A sufficiently rapid change in the electric tidal moment $G_{ij} = \partial_{ij}U$, with $U$ the Newtonian potential, can excite f-modes on the surface of any compact object, e.g.~\cite{Steinhoff:2016rfi}. In the case of a spin-aligned binary, this will generate f-modes with $m=0$ and $m=\pm2$. In this case, the amplitude of the f-modes are 2PN order, i.e. they scale like ${\cal{O}}(v^{4})$ with $v$ the orbital velocity. Thus, these effects are potentially subdominant compared to an intrinsic spheroidness, and we may once again assume $\epsilon_{2} \ll 1$.

Finally, a further example are deformed BHs in modified gravity, where uniqueness and no-hair theorems might not hold.\footnote{Although such deformed solutions exist (e.g.,~\cite{Cardoso:2018ptl,Sennett:2019bpc}) they arise from modified field equations that also affect the binary dynamics in other ways (e.g. by extra dissipative terms). Since we assume GR, our approach can describe this situation only if beyond-GR effects to the dynamics (e.g. modified fluxes) are negligible compared to the multipolar deformations.}

To be as general as possible while still working in the realm of astrophysical plausibility, we take $0 \le \epsilon_{2} <1$ for the remainder of this calculation. Under this assumption, the solution to Eq.~\eqref{eq:dVdpsi} is
\begin{equation}
    \label{eq:Vsol}
    {\cal{V}} = \tan^{-1}\left[\sqrt{\frac{1+\epsilon_{2}}{1-\epsilon_{2}}}\tan\psi_{2}\right]\,.
\end{equation}
Such an expression should be familiar to anyone who has studied Keplerian orbits, since it takes the same form as the mapping between the true anomaly $V$ and the eccentric anomaly $u$ for eccentric binaries, specifically
\begin{equation}
    \frac{V}{2} = \tan^{-1}\left[\sqrt{\frac{1+e}{1-e}}\tan\left(\frac{u}{2}\right)\right]\,.
\end{equation}
These expressions have known issues with branch cuts when $[\psi_{2}, u/2] = n\pi/2$ with $n$ an integer. However, it has been shown~\cite{Konigsdorffer:2006zt} through trigonometric identities that an equivalent expression that removes the branch cuts and properly tracks the secular behavior of $V$ with increasing $u$ is
\begin{equation}
    V = u + 2 \tan^{-1}\left(\frac{\beta_{e} \sin u}{1-\beta_{e} \cos u}\right)\,,
\end{equation}
where
\begin{equation}
    \beta_{e} = \frac{1 - \sqrt{1 - e^{2}}}{e}\,.
\end{equation}
Thus, an equivalent expression for ${\cal{V}}(\psi_{2})$ can be found by taking $V \rightarrow 2{\cal{V}}$ and $u \rightarrow 2\psi_{2}$, specifically
\begin{equation}
    \label{eq:V-polar}
    {\cal{V}} = \psi_{2} + \tan^{-1}\left[\frac{\beta_{2} \sin(2\psi_{2})}{1 - \beta_{2} \cos(2\psi_{2})}\right]\,,
\end{equation}
where $\beta_{2} = \beta_{e}(e\rightarrow \epsilon_{2})$.

Moving on to the inclination angle, we proceed by dividing Eq.~\eqref{eq:di-Q} by Eq.~\eqref{eq:psip}, and then divide by Eq.~\eqref{eq:dVdpsi}, to obtain
\begin{equation}
    \frac{d\iota}{d{\cal{V}}} = \frac{\epsilon_{2}}{\sqrt{1 - \epsilon_{2}^{2}}} \left[\frac{\tan\iota \sin(2{\cal{V}})}{1 + \epsilon_{2} \cos(2{\cal{V}})}\right]\,.
\end{equation}
Such an equation can be directly integrated to obtain
\begin{equation}
    \frac{\sin\iota}{\sin\iota_{0}} = \sqrt{\frac{1 + \epsilon_{2}}{1 + \epsilon_{2} \cos(2{\cal{V}})}}\,,
\end{equation}
where $\iota_{0}$ is the initial value of the inclination angle, i.e. $\iota_{0} = \iota({\cal V}=0)$. Using Eq.~\eqref{eq:Vsol}, this can be re-written in terms of $\psi_{2}$ as the dependent variable, specifically
\begin{equation}
    \label{eq:i-polar}
    \frac{\sin\iota}{\sin\iota_{0}} = \sqrt{\frac{1 + \epsilon_{2} \cos(2\psi_{2})}{1 + \epsilon_{2}}}\,.
\end{equation}
Note that in the limit $\iota_{0} \rightarrow 0$, the orbital angular momentum vector becomes aligned with the $Z$-axis of the body frame and $\iota$ becomes a constant. Thus, in the limit of alignment, there is no nutation.

Finally, moving on to the longitude of pericenter, we obtain an equation for $d\omega/d\psi_{2}$ by dividing Eq.~\eqref{eq:dw-Q} by Eq.~\eqref{eq:psip}. After some manipulation, this equation takes the form
\begin{equation}
    \frac{d\omega}{d\psi_{2}} = - \frac{c_{1} + c_{2} \cos(2\psi_{2})}{[1 - \epsilon_{2} \cos(2\psi_{2})] \sqrt{a_{-} - b \cos(2\psi_{2})}}\,,
\end{equation}
where
\begin{align}
    \label{eq:acoeff}
    a_{\pm} &= 1 \pm \epsilon_{2} - \sin^{2}\iota_{0}\,,
    \\
    b &= - \epsilon_{2} \sin^{2}\iota_{0}\,,
    \\
    c_{1} &= 3 - \epsilon_{2} \left(5 + 4\epsilon_{2}\right) + 5 \left(1 + \epsilon_{2}\right) \cos(2\iota_{0})\,,
    \\
    c_{2} &= \epsilon_{2} \left(1 + 5 \epsilon_{2}\right) - 5 \epsilon_{2} \left(1 + \epsilon_{2}\right) \cos(2\iota_{0})\,.
\end{align}
Naturally, this equation can be directly integrated to obtain
\begin{align}
    \label{eq:w-polar}
    \omega - \omega_{0} &= \frac{\sec\iota_{0}}{4 \sqrt{1 - \epsilon_{2}^{2}}} \left[\frac{c_{2}}{\epsilon_{2}} {\rm  EllF}\left(\psi_{2} \Big| \frac{2b}{b-a_{-}}\right) 
    \right.
    \nonumber \\
    &\left.
    - 4 (1 + \epsilon_{2}) {\rm Ell}\Pi\left(\frac{2\epsilon_{2}}{1 - \epsilon_{2}}; \psi_{2} \Big| \frac{2b}{b-a_{-}}\right)\right]\,,
\end{align}
where $\omega_{0}$ is the initial value, ${\rm EllF}$ and ${\rm Ell}\Pi$ are the elliptic integrals of the first and third kind, respectively. Note that this equation is divergent in the limit $\iota_{0} \rightarrow \pi/2$, since $\omega$ becomes ill-defined in this limit.

We leave the calculation of the solution to Eq.~\eqref{eq:psip} to the discussion in Sec.~\ref{sec:rr}.

\subsubsection{Axial Case}
\label{sec:axial}

The axial case is defined as the situation when $Q_{\pm 2}$ are zero, while the $Q_{\pm 1}$ coefficients are non-zero, which correspond to spherical harmonic modes that are odd under parity. Much of the setup for this case is the same as the polar case. We define the dimensionless parameters $\mathfrak{r}_{1} = Q_{+1}^{R}/Q_{0}$ and $\mathfrak{i}_{1} = Q_{+1}^{I}/Q_{0}$, from which we can define the \textit{axial modulus} and \textit{axial argument},
\begin{equation}
    \label{eq:a-params}
    \epsilon_{1} = \left[\frac{2}{3}\left(\mathfrak{r}_{1}^{2} + \mathfrak{i}_{1}^{2}\right)\right]^{1/2}\,, \qquad \alpha_{1} = \tan^{-1}\left(\mathfrak{i}_{1}/\mathfrak{r}_{1}\right)\,.
\end{equation}
We also modify the definition of $\psi_{2}$ to obtain $\psi_{1}$, specifically
\begin{equation}
    \frac{d\psi_{1}}{dt} = 3 \left(\frac{\pi}{5}\right)^{1/2} \frac{Q_{0}(1-e^{2})^{3/2}}{\nu M^{1/2} p^{7/2}} \sqrt{1 - \epsilon_{1}^{2}} \cos\iota\,.
\end{equation}
Lastly, we define ${\cal{V}} = \Omega + \alpha_{1}$. With these new variables, the relevant equations become
\begin{align}
    \label{eq:di-axial}
    \frac{d\iota}{d\psi_{1}} &= - \frac{\epsilon_{1} \cos {\cal{V}}}{\sqrt{1-\epsilon_{1}^{2}}}\,,
    \\
    \label{eq:dV-axial}
    \frac{d{\cal{V}}}{d\psi_{1}} &= \frac{1 + 2\epsilon_{1} \cot(2\iota) \sin {\cal{V}}}{\sqrt{1-\epsilon_{1}^{2}}}\,,
    \\
    \label{eq:dw-axial}
    \frac{d\omega}{d\psi_{1}} &= \frac{-5 \cos\iota + \sec \iota + \epsilon_{1} \left[3 - 5 \cot(2\iota)\right] \csc \iota \sin {\cal{V}}}{2\sqrt{1-\epsilon_{1}^{2}}} \,.
\end{align}
Unlike the polar case, the evolution of ${\cal{V}}$ is not decoupled from the evolution of $\iota$. It makes sense then to divide Eq.~\eqref{eq:di-axial} by Eq.~\eqref{eq:dV-axial} to obtain $d\iota/d{\cal{V}}$. Further, we make the change of variables $\gamma = \cot \iota$, which gives
\begin{equation}
    \frac{d\gamma}{d{\cal{V}}} = \frac{\gamma (1 + \gamma^{2}) \epsilon_{1} \cos {\cal{V}}}{\gamma - (1 - \gamma^{2}) \epsilon_{1} \sin {\cal{V}}}\,.
\end{equation}
This equation has a known exact solution, specifically
\begin{equation}
    \gamma = (1 + \gamma_{0}^{2}) \epsilon_{1} \sin {\cal{V}} + \sqrt{\gamma_{0}^{2} + (1 + \gamma_{0}^{2})^{2} \epsilon_{1}^{2} \sin^{2} {\cal{V}}}\,,
\end{equation}
where $\gamma_{0} = \cot\iota_{0}$ with $\iota_{0} = \iota({\cal V}=0)$.

With the solution for $\iota$ in hand, one can insert this into Eqs.~\eqref{eq:dV-axial}-\eqref{eq:dw-axial} and try to solve for $[{\cal{V}},\omega]$. Unfortunately, there does not appear to be a closed form solution to these for arbitrary value of $\epsilon_{1}$, even if we enforce the condition $\epsilon_{1} < 1$. We instead solve the equations perturbatively in $\epsilon_{1} \ll 1$, which is the case of most relevance to astrophysical scenarios. A straightforward calculation gives
\begin{align}
    {\cal{V}}(\psi_{1}) &= \psi_{1} + \sum_{n=1}^{\infty} \epsilon_{1}^{n} {\cal{V}}_{(n)}(\psi_{1})\,,
    \\
    \omega(\psi_{1}) - \omega_{0} &= \frac{1-4 \gamma_{0}^{2}}{2\gamma_{0} \sqrt{1 + \gamma_{0}^{2}}} \psi_{1} + \sum_{n=1}^{\infty} \epsilon_{1}^{n} \omega_{(n)}(\psi_{1})\,,
\end{align}
with $\omega_{0}$ and integration constant, and the first few functions in each sum given below,
\begin{align}
    {\cal{V}}_{(1)} &= - \frac{2}{\gamma_{0}} (1 - \gamma_{0}^{2}) \sin^{2}(\psi_{1}/2)\,,
    \\
    {\cal{V}}_{(2)} &= \frac{5}{2} \psi_{1} - \frac{(1-\gamma_{0}^{2})}{\gamma_{0}^{2}} \sin\psi_{1} - \frac{(1+\gamma_{0}^{4})}{2\gamma_{0}^{2}} \sin(2\psi_{1})\,,
    \\
    \omega_{(1)} &= \frac{(1 - 2\gamma_{0}^{2} + 2 \gamma_{0}^{4})}{2\gamma_{0}^{2} \sqrt{1 + \gamma_{0}^{2}}} \cos \psi_{1}\,,
    \\
    \omega_{(2)} &= - \frac{(1 - 6\gamma_{0}^{2} + 28 \gamma_{0}^{4})}{8 \gamma_{0}^{3} \sqrt{1 + \gamma_{0}^{2}}} \psi_{1}
    \nonumber \\
    &+ \frac{(1 - 3\gamma_{0}^{2} - 4\gamma_{0}^{4} + 2 \gamma_{0}^{6})}{2\gamma_{0}^{3}\sqrt{1 + \gamma_{0}^{2}}} \sin\psi_{1}
    \nonumber \\
    &- \frac{(3-8\gamma_{0}^{2} -4 \gamma_{0}^{4} - 8 \gamma_{0}^{6}}{16 \gamma_{0}^{3}\sqrt{1 + \gamma_{0}^{2}}} \sin(2\psi_{1})\,.
\end{align}
Note that the above solutions properly reduce to Eq.~\eqref{eq:obl-sol} in the limit $\epsilon_{1} \rightarrow 0$.

\subsubsection{Toward a General Solution}
\label{sec:generic}

Having considered the scenarios where analytic solutions are possible, some of which are in closed form, we may now work toward constructing general solutions to Eqs.~\eqref{eq:di-Q}-\eqref{eq:dw-Q}. We showed in Sec.~\ref{sec:polar} that the case with $\epsilon_{1} = 0$ and $\epsilon_{2} \ne 0$ admits closed form solutions. We choose to study the construction of a general solution by starting with the closed form solutions of Sec.~\ref{sec:polar} and consider the axial effects as a perturbation. The ansatz for the general solution will be
\allowdisplaybreaks[4]
\begin{align}
    \label{eq:i-full}
    \sin \iota &= \sin\left[\iota_{2}(\psi_{2})\right] + \sum_{n=1}^{\infty} \epsilon_{1}^{n} I_{(n)}(\psi_{2})\,,
    \\
    \Omega + \alpha_{2} &= {\cal{V}}_{2}(\psi_{2}) + \sum_{n=1}^{\infty} \epsilon_{1}^{n} V_{(n)}(\psi_{2})\,,
    \\
    \omega - \omega_{0} &= \omega_{2}(\psi_{2}) + \sum_{n=1}^{\infty} \epsilon_{1}^{n} W_{(n)}(\psi_{2})\,,
\end{align}
where $[\Omega_{0},\omega_{0}]$ are integration constants, and $\sin\iota_{2}$, ${\cal{V}}_{2}$, and $\omega_{2}$ are given as functions of $\psi_{2}$ in Eqs.~\eqref{eq:i-polar},\eqref{eq:V-polar}, and~\eqref{eq:w-polar}, respectively. To obtain the relevant equations for $\Lambda_{(n)}^{a}=[I_{(n)}, V_{(n)}, W_{(n)}]$, we insert the above ansatz into Eq.~\eqref{eq:di-Q}-\eqref{eq:dw-Q}, and expand about $\epsilon_{1} \ll 1$. 

To order ${\cal{O}}(\epsilon_{1}^{0})$, the osculating equations are automatically satisfied. To higher order, we obtain equations of the schematic form
\begin{align}
    \frac{d\Lambda^{a}_{(n)}}{d\psi_{2}} &= F_{(n)}^{a}\left[\psi_{2}, \Lambda_{(1)}^{b}(\psi_{2}), ..., \Lambda_{(n-1)}^{b}(\psi_{2});\epsilon_{2}\right]\,.
\end{align}
In practice, we have failed to find closed form solutions to these equations for arbitrary $\epsilon_{2}$, and have instead sought to solve them in the limit $\epsilon_{2}\ll1$. The solutions take the general form of a power series, specifically
\begin{equation}
    \Lambda_{(n)}^{a} = \sum_{k=0}^{\infty} \epsilon_{2}^{k} \Lambda_{(n,k)}^{a}(\psi_{2})\,.
\end{equation}
Defining $\Delta = \alpha_{1} - \alpha_{2}$, the solutions up to order ${\cal{O}}(\epsilon_{1} \epsilon_{2})$ are
\allowdisplaybreaks[4]
\begin{widetext}
\begin{align}
    I_{(1,0)} &= -2 \cos\iota_{0} \cos\left(\Delta + \frac{\psi_{2}}{2}\right)\sin\left(\frac{\psi_{2}}{2}\right)\,,
    \\
    I_{(1,1)} &= \frac{1}{6} \sec\iota_{0} \sin\left(\frac{\psi_{2}}{2}\right) \left\{6\cos\psi_{2} \sin\Delta - \cos(2\iota_{0}) \left[-6\sin\Delta + 9 \sin(\Delta-\psi_{2}) + 5 \sin(\Delta+\psi_{2}) + 10 \sin(\Delta +2\psi_{2})\right]\right\}\,,
    \\
    V_{(1,0)} &= -2 \cot(2\iota_{0}) \left[\cos\Delta - \cos(\Delta + \psi_{2})\right]\,,
    \\
    V_{(1,1)} &= \frac{1}{48} \big\{4 \left[9 + 10 \cos(2\iota_{0}) + 5 \cos(4\iota_{0})\right]\csc\iota_{0} \sec^{3}\iota_{0} \sin\Delta \sin^{3}\psi_{2} + \cos\Delta \sec\iota_{0}\left[-48 \cos(2\iota_{0} \cos(2\psi_{2}) \csc\iota_{0} \right.
    \nonumber \\
    &\left.
    + 3 \left(-7 + 10 \cos[2\iota_{0}] + 5 \cos[4\iota_{0}]\right) \cos\psi_{2} \csc\iota_{0} \sec^{2}\iota_{0} + \left(9 + 10 \cos[2\iota_{0}] + 5 \cos[4\iota_{0}]\right) \cos(3\psi_{2}) \csc\iota_{0} \sec^{2}\iota_{0} 
    \right.
    \nonumber \\
    &\left.
    + \left(2 + \cos[2\iota_{0}]\right) \sec\iota_{0} \tan\iota_{0}\right]\big\} \,,
    \\
    W_{(1,0)} &= \frac{1}{8} \sec\iota_{0} \tan\iota_{0} \Big\{2\psi_{2} \sin\Delta \big[7 + 5 \cos(2\iota_{0})\big] - \csc^{2}\iota_{0} \left[7 + 4 \cos(2\iota_{0}) + 5 \cos(4\iota_{0})\right] \sin\left(\Delta + \frac{\psi_{2}}{2}\right) \sin\left(\frac{\psi_{2}}{2}\right) \Big\}\,,
    \\
    W_{(1,1)} &= \frac{1}{576} \csc\iota_{0} \sec^{4}\iota_{0} \Big\{ \cos\Delta \Big[-2 \left(11 - 323 \cos[2\iota_{0}] + 65 \cos[4\iota_{0}] + 55 \cos[6\iota_{0}]\right) \nonumber \\
    &+ 36 \left(2 + \cos[2\iota_{0}]\right) \left(3 - 16 \cos[2\iota_{0}] + 5 \cos[4\iota_{0}]\right) \cos\psi_{2} + 9 \left(12 + 35 \cos[2\iota_{0}] + 12 \cos[4\iota_{0}] + 5 \cos[6\iota_{0}]\right) \cos(2\psi_{2}) 
    \nonumber \\
    &- \left(14 + 7 \cos[2\iota_{0}] + 50 \cos[4\iota_{0}] + 25 \cos[6\iota_{0}]\right)\cos(3\psi_{2})\Big] + \sin\Delta \Big[9 \left(-2 - 37 \cos[2\iota_{0}] + 2 \cos[4\iota_{0}] + 5 \cos[6\iota_{0}]\right) \sin\psi_{2} 
    \nonumber \\
    &+ 6 \sin^{2}\iota_{0} \left(\psi_{2} \left[131 + 188 \cos\{2\iota_{0}\} +65 \cos\{4\iota_{0}\}\right] + 3 [3 + \cos\{2\iota_{0}\}] [-1 + 5 \cos\{2\iota_{0}\}] \sin[2\psi_{2}]\right) 
    \nonumber \\
    &+ \left(14 + 7 \cos[2\iota_{0}] + 50 \cos[4\iota_{0}] + 25 \cos[6\iota_{0}]\right) \sin(4\psi_{2})\Big]\Big\}\,.
\end{align}
\end{widetext}
We stop the expansion here since the solutions for $\Lambda_{(n,k)}^{a}$ become increasingly complicated. The above discussion and results present a schematic of the calculation, and one can easily extend these results to higher order in $\epsilon_{1,2}$ if desired.

\begin{figure*}[htb!]
    \centering
    \includegraphics[width=\textwidth, trim={6cm, 2cm, 6cm, 2cm}, clip]{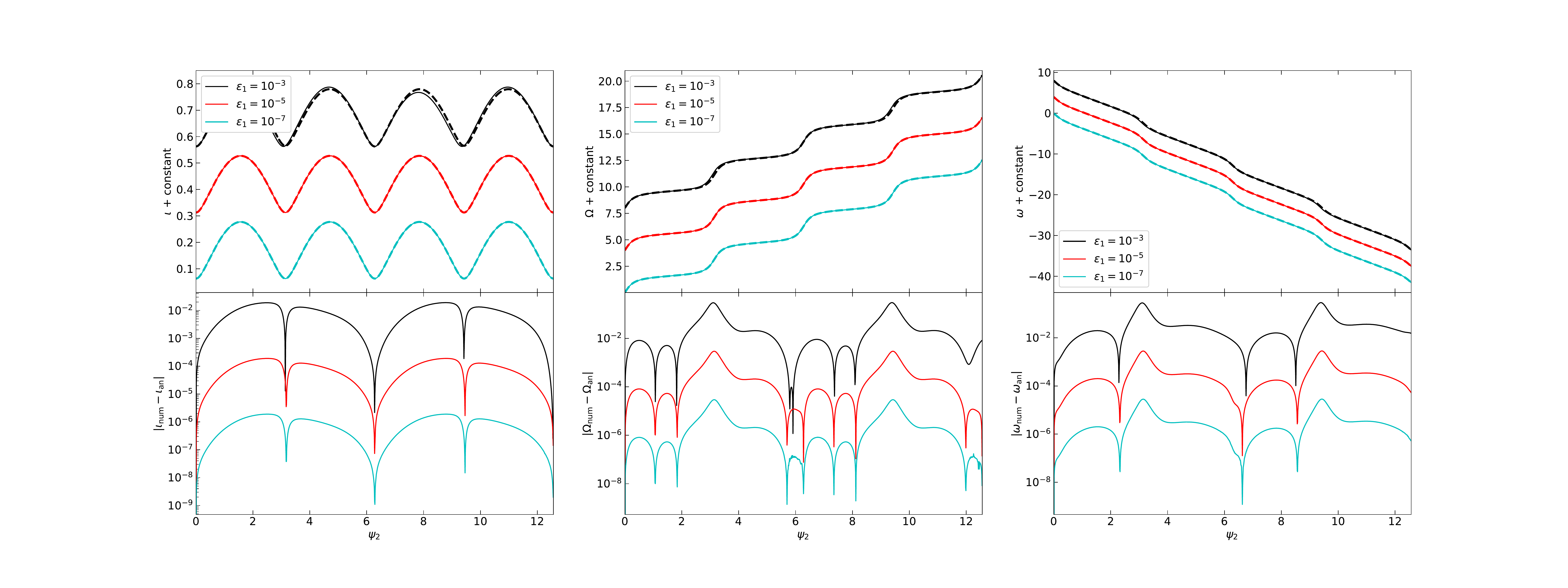}
    \caption{Top Panel: Comparison of the numerical evolution of $\iota$ (left), $\Omega$ (middle), and $\omega$ (right) from Eqs.~\eqref{eq:di-Q}-\eqref{eq:dw-Q} (solid lines) to the analytic solutions (dashed lines) given in Sec.~\ref{sec:generic} for $\epsilon_{2}=0.9$, $\alpha_{2}=0$, $\alpha_{1} = \pi/2$. Each color represents a different value of $\epsilon_{1}$, specifically $10^{-3}$(black/upper), $10^{-5}$(red/middle), and $10^{-7}$(cyan/lower). All of the calculations are performed with $\iota_{0} = \pi/50$. Each case is offset by a constant value from the others so that they do no overlap. Bottom Panels: Relative difference between the numeric and analytic solutions.}
    \label{fig:prec_comp}
\end{figure*}

To provide an estimate of the accuracy of the above solutions, we compare these analytic results to numerical evolutions of Eqs.~\eqref{eq:di-Q}-\eqref{eq:dw-Q}. In order to perform the numerical integration of the equations, we convert them to equations of the form $d\mu^{a}/d\psi_{2}$, and recast them in terms of $\epsilon_{1,2}$ and $\alpha_{1,2}$ instead of the $Q_{m}$ coefficients. We set $\epsilon_{2}=0.9$, $\alpha_{2}=0$, $\alpha_{1} = \pi/2$, and study the behavior of the solutions for different values of $\epsilon_{1}$, specifically $\epsilon_{1} = [10^{-7}, 10^{-5}, 10^{-3}]$. The results of this comparison are shown in Fig.~\ref{fig:prec_comp}. The top panels show the plots of both the numeric (solid lines) and analytic (dashed lines) for (from left to right) $\iota$, $\Omega$, and $\omega$. The bottom panels display the relative difference between the two solutions, showing the error introduced in each phase. As $\epsilon_{1}$ decreases by each order of magnitude, so does the error by a comparable amount. Note that this has to be the case since the analytic solutions are exact in the limit $\epsilon_{1} \rightarrow 0$.

\subsection{Non-precessing Solutions}
\label{align}

Alignment between the $Z$-axis of the body frame and $\hat{L}$, specifically $\iota = 0$ and $\langle d\iota/dt \rangle = 0$, is only possible when $Q_{\pm 1} = 0$. When this occurs, the orbital basis of Eqs.~\eqref{eq:n-vec}-\eqref{eq:L-vec} reduces down to
\begin{align}
    \hat{n} &= \left[\cos(\phi + \Omega), \sin(\phi + \Omega), 0\right]\,,
    \\
    \hat{\lambda} &= \left[-\sin(\phi + \Omega), \cos(\phi + \Omega), 0\right]\,,
    \\
    \hat{L} &= \left[0, 0, 1\right]\,,
\end{align}
and thus the effect of the mass quadrupole contributions is to modulate the orbital phase $\phi$ by $\Omega$. 

The solutions in this limit can be considered as a special class of the polar solutions. Defining $[\psi_{2}, \epsilon_{2}, \alpha_{2}]$ as in Eqs.~\eqref{eq:p-params} and~\eqref{eq:psip}, it follows that ${\cal{V}} = \Omega + \alpha_{2}$ still obeys Eq.~\eqref{eq:V-polar}. The longitude of pericenter reduces to
\begin{align}
    \omega-\omega_{0} &= -\frac{1}{\sqrt{1-\epsilon_{2}^{2}}} \left[\psi_{2} + (1+\epsilon_{2}) \Pi\left(\frac{2\epsilon_{2}}{1-\epsilon_{2}}; \psi_{2} \Big| 0\right) \right]\,.
\end{align}
A straightforward calculation shows that the direction of pericenter advances at the constant rate
\begin{equation}
    \lim_{\iota\rightarrow0} \left(\frac{d\omega}{d\psi_{2}} + \cos\iota \frac{d\Omega}{d\psi_{2}}\right) = -\frac{1}{\sqrt{1-\epsilon_{2}^{2}}}\,.
\end{equation}
The motion in the aligned limit is now fully specified.

\subsection{Oscillatory Effects \& Orbital Motion}
\label{sec:oscorb}

As mentioned at the beginning of this section, $\langle de/dt \rangle = 0 = \langle dp/dt \rangle$, and thus the leading order correction to $[e,p]$ comes from the oscillatory effects of Eq.~\eqref{eq:msa-osc}. In order to complete our solution to the dynamics of the binary to leading order in the mass quadrupole moment (or alternatively, complete to 2PN order), we must also consider these oscillatory effects in $[e,p]$. We do not need to consider these corrections to the angles $[\iota, \Omega, \omega]$, since these terms do not enter relevant orbital quantities until relative 2PN order, and thus their oscillatory effects will be further suppressed to relative 4PN order. Following Eq.~\eqref{eq:msa-osc}, the solutions for $[e,p]$ are schematically
\begin{align}
    \label{eq:e-osc}
    e(V) &= e_{0} + \frac{1}{p_{0}^{2}} \sum_{k=0}\left[ {\cal{C}}_{k}^{e} \cos(kV) + {\cal{S}}_{k}^{e} \sin(kV)\right] + {\cal{O}}(p_{0}^{-4})\,,
    \\
    \label{eq:p-osc}
    \frac{p(V)}{p_{0}} &= 1 + \frac{1}{p_{0}^{2}} \sum_{k} \left[ {\cal{C}}_{k}^{p} \cos(kV) + {\cal{S}}_{k}^{p} \sin(kV) + {\cal{O}}(p_{0}^{-4})\right]\,,
\end{align}
where $[e_{0}, p_{0}]$ are determined by initial conditions. We do not provide the specific $[{\cal{C}}_{k}, {\cal{S}}_{k}]$ coefficients in this work, since they are rather long and unenlightening. However, they can be readily computed from Eqs.~\eqref{eq:dp-osc}-\eqref{eq:de-osc}.

Naively, one might expect that $e_{0}$ would correspond to the actual (in a geometric sense) eccentricity of the orbit, and likewise for the semi-latus rectum $p_{0}$. However, this is not in general true. Indeed, one can easily check that $e_{0} = 0$ does not correspond to circular orbits, since $\lim_{e_{0}\rightarrow0}e(V) \ne 0$. We, thus, have to redefine these parameters such that they make physical sense. To do this, we follow~\cite{Mora:2003wt}, and define the new eccentricity and semi-latus rectum as
\begin{align}
    \tilde{e} &= \frac{\sqrt{\Omega_{p}} - \sqrt{\Omega_{a}}}{\sqrt{\Omega_{p}} + \sqrt{\Omega_{a}}}\,,
    \\
    \tilde{p} &= M \left(\frac{2}{\sqrt{M \Omega_{p}} + \sqrt{M \Omega_{a}}}\right)^{4/3}\,,
\end{align}
where $[\Omega_{p}, \Omega_{a}] = [\max dV/dt, \min dV/dt]$. The resulting expressions are
\begin{align}
    \label{eq:e-new}
    \tilde{e} &= e_{0} + \sqrt{\frac{\pi}{5}} \frac{1}{8 \nu M p_{0}^{2}} \sum_{a} Q^{a} \tilde{\cal{E}}_{a}(\iota,\Omega,\omega) + {\cal{O}}(p_{0}^{-4})\,,
    \\
    \label{eq:p-new}
    \tilde{p} &= p_{0} \left[1 + \sqrt{\frac{\pi}{5}} \frac{1}{3 \nu M p_{0}^{2}} \sum_{a} Q^{a} \tilde{\cal{P}}_{a}(\iota,\Omega,\omega) + {\cal{O}}(p_{0}^{-4})\right]\,,
\end{align}
where $Q^{a} = [Q_{0}, Q^{R,I}_{1}, Q^{R,I}_{2}]$, and the coefficients $[\tilde{\cal{E}}_{a}, \tilde{\cal{P}}_{a}]$ are given in Appendix~\ref{app:coeffs}. This allows us to now properly take the circular limit $\tilde{e} \rightarrow 0$, since that is the limit of most relevance to ground based GW detectors. For the remainder of the calculation, we work in this limit.

In the next section, we will calculate the leading order effects due to radiation reaction on the binary, but in order to complete that, we need two more quantities from the orbital dynamics, namely the modifications to Kepler's third law and the on-shell orbital energy. The former of these allows us to relate the orbital velocity to the orbital frequency. To do so, we begin by calculating the corrections to the orbital period, which is computed by
\begin{equation}
    T_{\rm orb} = \int_{0}^{2\pi} \left(\frac{dV}{dt}\right)^{-1} dV\,.
\end{equation}
The orbital frequency is then $F = 1/T_{\rm orb}$. With the orbital velocity $v = \tilde{p} \dot{\phi}$ in the limit $\tilde{e} = 0$, we obtain
\begin{equation}
    \label{eq:kep3}
    v = \tilde{u} \left[1 + \frac{\tilde{u}^{4}}{72 \nu M^{3}} \sqrt{\frac{\pi}{5}} \sum_{a} Q^{a} \tilde{\Omega}_{a}(\iota,\Omega,\omega) + {\cal{O}}(\tilde{u}^{8})\right]\,,
\end{equation}
where $\tilde{u} = (2\pi M F)^{1/3}$ and the coefficients $\tilde{\Omega}_{a}$ are given in Appendix~\ref{app:coeffs}.

Lastly, we need the on-shell orbital energy in order to utilize the balance law for radiation reaction. The starting point is Eq.~\eqref{eq:Eorb}, including only the Newtonian and quadrupole terms. To evaluate this on shell, one has to insert Eqs.~\eqref{eq:r-kep}-\eqref{eq:phidot-kep} into this to write $E_{\rm orb}$ in terms of the osculating quantities $\mu^{a}$. This expression is still dependent on the orbital timescale through $V$. To address this, we then have to combine this expression with Eqs.~\eqref{eq:e-osc}-\eqref{eq:p-osc} and truncate at the relevant PN order. The final result, which is independent of $V$, is
\begin{equation}
    \label{eq:E-os}
    E_{\rm orb} = -\frac{1}{2} \mu \tilde{u}^{2} \left[1 + \frac{\tilde{u}^{4}}{36 \nu M^{3}} \sqrt{\frac{\pi}{5}} \sum_{a} Q^{a} \tilde{E}_{a}(\iota,\Omega,\omega) + {\cal{O}}(\tilde{u}^{8})\right]
\end{equation}
where we have made use of Eqs.~\eqref{eq:e-new}-\eqref{eq:p-new}, and taken the limit $\tilde{e}\rightarrow0$. The coefficients $\tilde{E}_{a}$ are also given in Appendix~\ref{app:coeffs}.

It is worth noting that all of the orbital quantities derived in this section vary on the precessing timescale through $[\iota,\Omega,\omega]$. We discuss in detail how to handle this behavior in the next section.
\section{GW Emission}
\label{sec:gws}

The solutions of the previous section constitute the solutions to the conservative dynamics of the binary in the presence of generic mass quadrupole effects. In this section, we will consider the effects of dissipation on such systems through the emission of GWs. 

\subsection{Radiation reaction}
\label{sec:rr}

We wish to compute the leading PN order corrections to the inspiral of compact binaries due to generic quadrupole effects. In order to do so, it suffices to consider the leading PN order effects in radiation reaction, which are governed by the quadrupole approximation. The energy ${\cal{P}}$ and angular momentum ${\cal{J}}^{i}$ fluxes due to GWs therein are governed by
\begin{align}
    \label{eq:GW-flux}
    {\cal{P}} &= \frac{1}{5c^{5}} \dddot{I}^{<ij>} \dddot{I}^{<ij>}\,,
    \\
    {\cal{J}}^{i} &= \frac{2}{5c^{5}} \epsilon^{ijk} \ddot{I}^{<jq>} \dddot{I}^{<kq>}\,,
\end{align}
with $I^{ij}$ the orbital quadrupole moment of the binary. The quadrupole deformation of the body does not explicitly contribute to these equations since we are assuming that the $Q_{m}$'s are static. However, they do contribute implicitly through the definition of $I^{ij}$, specifically
\begin{equation}
    \label{eq:orb-quad}
    I^{ij} = \mu x^{i} x^{j} + {\cal{O}}(c^{-2})\,,
\end{equation}
where $r$ is given by Eq.~\eqref{eq:r-kep} and $n^{i}$ is given by Eq.~\eqref{eq:n-vec}. Due to the osculating nature of the orbit, when taking the time derivatives of $I^{ij}$, we must act on the elements $[\iota,\omega,\Omega]$ in addition to the orbital phase $\phi$. For each time derivative acting on the former, we are required to insert the osculating equations in Eqs.~\eqref{eq:di-osc}-\eqref{eq:dw-osc}, and then accurately PN truncate them.

In the present calculation, we take the limit $\tilde{e}=0$, since most binaries of relevance to ground based detectors will have negligible eccentricity. We then calculate the rate of change of the orbital energy, which is related to the energy flux through the balance law
\begin{equation}
    \label{eq:balaw}
    \frac{dE_{\rm orb}}{dt} = - \langle {\cal{P}} \rangle\,,
\end{equation}
and where $E_{\rm orb}$ is given by Eq.~\eqref{eq:Eorb}. Since we only desire the leading PN order correction, it suffices to work at relative Newtonian order, meaning we only need to consider the contributions $E_{\rm N}$ and $E_{\rm quad}$ when using Eq.~\eqref{eq:Eorb}. 

We must begin by evaluating Eq.~\eqref{eq:GW-flux}. The time derivative can be performed by acting on Eq.~\eqref{eq:orb-quad} directly. The one important feature of said procedure is that every time an instance of the acceleration $a^{i} = \ddot{x}^{i}$ appears, we must insert the equations of motion in Eq.~\eqref{eq:orbitaleom}. Since we are working to relative Newtonian order, it suffices to only consider the Newtonian and mass quadrupole terms in the relative acceleration equation. Doing so, we end up with $\dddot{I}^{ij} = \dddot{I}_{0}^{ij} + \delta (\dddot{I}^{ij})$, where
\begin{align}
    \label{eq:dIorb}
    \dddot{I}^{ij}_{0} &= \mu M \left[\frac{6\dot{r}}{r^{4}} x^{i} x^{j} - \frac{8}{r^{3}} v^{(i} x^{j)}\right]\,,
    \\
    \delta \left(\dddot{I}^{ij}\right) &= 2\mu \left[2 f_{\rm pert}^{(i} v^{j)} + \dot{f}_{\rm pert}^{(i} x^{j)}\right]\,,
\end{align}
where $f_{\rm pert}^{i}$ is given by Eq.~\eqref{eq:quadmass}. The energy flux can now be directly computed using Eqs.~\eqref{eq:r-kep}-\eqref{eq:L-vec}. When taking the orbit average necessary for the balance law, we must also take into account the corrections to the orbital period and $dV/dt$, both of which are detailed in Sec.~\ref{sec:oscorb}.

On the other hand, the orbital energy is given in Eq.~\eqref{eq:E-os}. When taking a time derivative of this expression, one has to remember that now $\tilde{u}$ is being promoted to a function of time. The orbital energy also depends on $\mu^{a}$, and one would need to insert the osculating equations $d\mu^{a}/dt$ everywhere these terms appear. However, we are only working to relative Newtonian order, so these will introduce higher PN order effects that we may neglect. Thus, we are left with only terms depending on $d\tilde{u}/dt$ in Eq.~\eqref{eq:balaw}. Solving, we obtain
\begin{equation}
    \label{eq:dudt}
    \frac{d\tilde{u}}{dt} = \frac{32}{5} \frac{\nu}{M} \tilde{u}^{9} \left[1 + \frac{\tilde{u}^{4}}{8 M^{3} \nu} \sqrt{\frac{\pi}{5}} \sum_{a} Q^{a} \tilde{U}_{a}(\iota,\Omega,\omega)\right]
\end{equation}
where the coefficients $\tilde{U}_{a}$ are given in Appendix~\ref{app:coeffs}. We now have the necessary equation to solve for the evolution of the binary under radiation reaction.

In general, radiation reaction introduces a new timescale to the problem, in addition to the orbital timescale encoded in $V$ and the precession timescale encoded in $\psi_{2}$. To consistently solve the problem, one has to, once again, solve the relevant equations in a multiple scale analysis, now with three timescales. As we detailed back in Sec.~\ref{sec:osculat}, the leading order behavior is obtained by averaging over the relevant oscillatory scales. In the process of deriving Eq.~\eqref{eq:dudt}, we already performed the average over the oscillatory orbital timescale. Observe that the coefficients $\tilde{U}_{a}$ only depend on $\psi_{2}$ through oscillatory functions of $[\iota,\Omega,\omega]$. Thus, the precession effects in Eq.~\eqref{eq:dudt} are actually oscillatory effects, and it suffices to perform a precession average, i.e.
\begin{equation}
    \langle f \rangle_{\psi_{2}} = \frac{1}{2\pi} \int_{0}^{2\pi} f(\psi_{2}) d\psi_{2}\,.
\end{equation}
We will thus obtain double averaged equations that constitute the leading order behavior under radiation reaction. The precession average of Eq.~\eqref{eq:dudt} is found by simply making the replacement $U_{a} \rightarrow \langle U_{a} \rangle_{\psi_{2}}$. In general, the averages do not admit closed form expressions for arbitrary $\epsilon_{1,2}$. Given the astrophysical considerations discussed in Sec.~\ref{sec:prec}, we compute these in an expansion $\epsilon_{1} \ll 1 \gg \epsilon_{2}$, which provides us the mapping
\begin{equation}
    \label{eq:avg-coeffs}
    \sum_{a} Q^{a} \langle \tilde{U}_{a} \rangle_{\psi_{2}} \rightarrow Q_{0} \sum_{pq} \epsilon_{1}^{p} \epsilon_{2}^{q} \; {\cal{U}}_{pq}(\iota_{0}, \omega_{0}, \alpha_{1}, \alpha_{2})
\end{equation}
where the coefficients only depend on constants of the precession dynamics and are listed in Appendix~\ref{app:coeffs}. For brevity, we will drop the explicit sum over $[p,q]$ and apply the Einstein summation convention in future expressions.

In order to compute the Fourier domain gravitational waveform through the SPA, we require three phases, namely $[t(\tilde{u}),\phi(\tilde{u}),\psi_{2}(\tilde{u})]$. The first of these is found by inverting $\langle d\tilde{u}/dt \rangle_{\psi_{2}}$, specifically
\begin{align}
    \label{eq:tofu}
    t(\tilde{u}) &= t_{c} + \int d\tilde{u} \left(\Big\langle \frac{d\tilde{u}}{dt} \Big\rangle_{\psi_{2}}\right)^{-1}
    \nonumber \\
    &= t_{c} - \frac{5 M}{256 \nu \tilde{u}^{8}} \left[1 - \frac{Q_{0}\tilde{u}^{4}}{4M^{3} \nu }\sqrt{\frac{\pi}{5}} \epsilon_{1}^{p} \epsilon_{2}^{q} \; {\cal{U}}_{pq} \right]
\end{align}
where $t_{c}$ is the time of coalescence. Similarly, the orbital phase is $d\phi/dt = \tilde{u}^{3}/M$, and thus
\begin{align}
    \label{eq:phiofu}
    \phi(\tilde{u}) &= \phi_{c} + M^{-1} \int d\tilde{u} \; \tilde{u}^{3} \left(\Big\langle \frac{d\tilde{u}}{dt} \Big\rangle_{\psi_{2}}\right)^{-1}
    \nonumber \\
    &= \phi_{c} - \frac{1}{32 \nu \tilde{u}^{5}} \left[1 - \frac{5Q_{0}\tilde{u}^{4}}{8 M^{3} \nu} \sqrt{\frac{\pi}{5}} \epsilon_{1}^{p} \epsilon_{2}^{q} \; {\cal{U}}_{pq}\right]
\end{align}
where $\phi_{c}$ is the phase of coalescence.

The evolution of the precession phase $\psi_{2}(\tilde{u})$ requires more careful consideration. The time evolution of $\psi_{2}$ is given in Eq.~\eqref{eq:psip}. Unlike the functions $[t(\tilde{u}), \phi(\tilde{u})]$ where we could average over $\psi_{2}$, we cannot do so here and must consider the full evolution equations $d\psi_{2}/d\tilde{u}$, which is obtained by dividing Eq.~\eqref{eq:psip} by Eq.~\eqref{eq:dudt}. However, since the evolution of $\psi_{2}$ is already of linear order in the mass quadrupole moments and of absolute 2PN order, it suffices for our purposes to truncate this expression to leading PN order, obtaining
\begin{equation}
    \frac{d\psi_{2}}{d\tilde{u}} = \frac{3\sqrt{5\pi}}{32} \frac{Q_{0} \sqrt{1-\epsilon_{2}^{2}}}{M^{2} \nu \tilde{u}^{2}} \cos\iota\,,
\end{equation}
where $\iota$ is a function of $\psi_{2}$ through Eq.~\eqref{eq:i-full}. We solve this perturbatively in $\epsilon_{1} \ll 1$, while keeping $\epsilon_{2}$ arbitrary, just as we did in Sec.~\ref{sec:generic}. Writing $\psi_{2}(\tilde{u}) = \psi_{2,0}(\tilde{u}) + \epsilon_{1} \psi_{2,1}(\tilde{u}) + {\cal{O}}(\epsilon_{1}^{2})$, we obtain to leading order
\begin{equation}
    \label{eq:psi20u}
    \frac{d\psi_{2,0}}{d\tilde{u}} = \frac{3\sqrt{5\pi}}{32} \frac{Q_{0} \sqrt{1-\epsilon_{2}}}{M^{2}\nu \tilde{u}^{2}} \left[a_{+} + b \cos(2\psi_{2})\right]^{1/2}\,.
\end{equation}
This expression can be directly integrated by moving all terms dependent on $\psi_{2}$ to the left hand side and integrating. The resulting integral on the left hand side produces the elliptic integral of the first kind ${\rm EllF}[\psi_{2} | 2b/(b+a_{+})]$. Despite the dependence on specialized functions, the resulting equality can be solved to obtain $\psi_{2,0}(\tilde{u})$ by utilizing the fact that the Jacobi amplitude function ${\text{am}}(x|n)$ is the inverse of the ellitpic integral of the first kind, specifically ${\rm am}[{\rm EllF}(x|n)|n] = x$. Rearranging, we obtain
\begin{equation}
    \psi_{2,0}(\tilde{u}) = {\rm am} \left[{\rm EllF}\left(\psi_{c} \Bigg| \frac{2b}{b+a_{+}}\right) - \Psi_{2}(\tilde{u}) \Bigg| \frac{2b}{b+a_{+}}\right]
\end{equation}
where $a_{+}$ is given in Eq.~\eqref{eq:acoeff}, and
\begin{equation}
    \Psi_{2}(\tilde{u}) = \frac{3\sqrt{5\pi}}{32} \frac{Q_{0} \sqrt{1-\epsilon_{2}^{2}}}{M^{2} \nu \tilde{u}} \cos\iota_{0}\,.
\end{equation}
This expression is exact in the limit $\epsilon_{1} \rightarrow 0$, and for $\epsilon_{2} \in [0,1)$. Note that $\Psi_{2} \sim \tilde{u}^{-1}$, and is thus a $-0.5$PN effect, unlike the orbital phase in Eq.~\eqref{eq:phiofu} which scales as $\tilde{u}^{-5}$ and enters at $-2.5$PN order.

The correction to the precession phase $\psi_{2,1}(\tilde{u})$ due to axial modes require a more in depth calculation. Similar to the results in Sec.~\ref{sec:generic}, there does not appear to be a closed form solution to this for arbitrary $\epsilon_{2}$, and we instead solve them in the limit $\epsilon_{2} \ll 1$. To leading order,
\begin{equation}
    \frac{d\psi_{2,1}}{d\tilde{u}} = \frac{3\sqrt{5\pi}}{16} \frac{Q_{0} \sin\iota_{0}}{M^{3} \nu} \left\{\frac{\sin\left[\Delta + \psi_{c} - \Psi_{0}(\tilde{u})\right]-\sin\Delta}{2u^{2}}\right\}\,.
\end{equation}
where $\Psi_{0} = \underset{\epsilon_{2}\rightarrow0}{\lim}\Psi_{2}$. This can be directly integrated to obtain
\begin{align}
    \label{eq:psi21u}
    \psi_{2,1}(\tilde{u}) &= \sin\Delta \tan\iota_{0} \Psi_{0}(\tilde{u}) 
    \nonumber \\
    &- \tan\iota_{0} \cos\left[\Delta + \psi_{c} - \Psi_{0}(\tilde{u})\right] + {\cal{O}}(\epsilon_{2})\,.
\end{align}
The calculation can be extended to include the ${\cal{O}}(\epsilon_{2}^{n})$ corrections to $\psi_{2,1}(\tilde{u})$ in a straightforward way. We do not calculate them here for brevity, as well as the fact that these terms will scale as $\epsilon_{1} \epsilon_{2}^{n}$ in the precession phase, and can thus be treated as higher order.

\begin{figure*}[htb!]
    \centering
    \includegraphics[width=\textwidth, trim={6cm, 2cm, 6cm, 2cm}, clip]{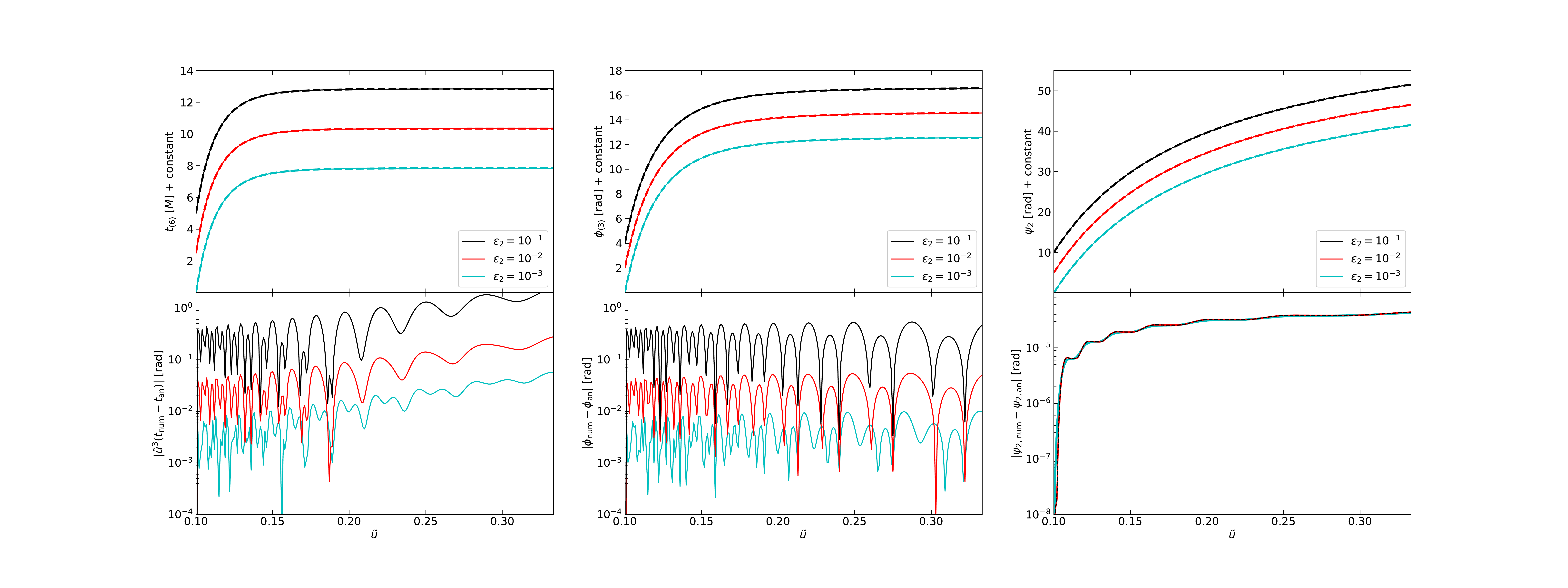}
    \caption{Top Panel: Comparison of the numerical evolution (solid lines) of the time variable $t_{(6)}(\tilde{u}) = t(\tilde{u})/10^{6}$ (left), orbital phase $\phi_{(3)}(\tilde{u}) = \phi(\tilde{u})/10^{3}$ (middle), and precession phase $\psi_{2}(\tilde{u})$ to the analytic PN expressions (dashed lines) in Eq.~\eqref{eq:tofu}, Eq.~\eqref{eq:phiofu}, and Eqs.~\eqref{eq:psi20u} \&~\eqref{eq:psi21u}, respectively. The dashed lines correspond to different values of the polar modulus $\epsilon_{2}$, specifically $10^{-3}$ (cyan), $10^{-2}$ (red), and $10^{-1}$ (black). The remaining parameters are held fixed at $\epsilon_{1} = 10^{-3}$, $\alpha_{2} = 0$, $\alpha_{1} =\pi/2$, $\nu =1/4$, and $Q_{0} = M^{3}$. Bottom Panel: Dephasing in radians between the numerical evolution and analytic expressions. The dephasing in $\psi_{2}$ (bottom right) does not vary significantly for the values of $\epsilon_{2}$ considered and it is mainly due to the PN truncation and the precession-average procedure.}
    \label{fig:rr_comp}
\end{figure*}

In Fig.~\ref{fig:rr_comp}, we compare the analytic approximation of $t(\tilde{u})$, $\phi(\tilde{u})$, $\psi_{2}(\tilde{u})$ derived in this section to numerical evolutions of the precessions equations in Eqs.~\eqref{eq:di-osc} coupled to Eqs.~\eqref{eq:dudt} to include radiation reaction. For these numerical evolutions, we fixed $\epsilon_{1} = 10^{-3}$ and varied $\epsilon_{2} = [10^{-3},10^{-2},10^{-1}]$. In doing so, we found that the dephasing between the two solutions, which encodes the error in the analytic expressions, depends on $\epsilon_{1}$ only mildly. For $t(\tilde{u})$ and $\phi(\tilde{u})$, the dephasing becomes of order one radian for the largest value of $\epsilon_{2}$. The dephasing for these quantities can be improved by carrying the $\epsilon_{1} \ll 1 \gg \epsilon_{2}$ expansion to higher order. On the other hand, the dephasing in $\psi_{2}(\tilde{u})$ does not vary significantly for varying $\epsilon_{2}$, since it can be solved for exactly in the case of polar configurations.

\subsection{Gravitational waveform}
\label{sec:sua}

Let us now consider the gravitational waveform of a binary with arbitrary mass quadrupole coefficients. For simplicity, we will seek to develop the corrections to the TaylorF2 waveforms for quasi-circular binaries due to generic mass quadrupoles. To derive this, it suffices to consider the quadrupole approximation, where the metric perturbation is given by
\begin{equation}
    h_{ij} = \frac{2}{c^4 D_{L}} \ddot{I}_{<ij>}\,,
\end{equation}
where $D_{L}$ is the luminosity distance to the source. The orbital quadrupole moment must be handled in the manner described above Eq.~\eqref{eq:dIorb} when working in the osculating formalism. The observable waveform is found by projecting $h_{ij}$ into the transverse trace-less (TT) gauge. In order to do this, we define the line of sight vector $N^{i} = [\sin\theta_{N} \cos\phi_{N}, \sin\theta_{N} \cos\phi_{N}, \cos\theta_{N}]$, where $\theta_{N}$ is the angle between the Z-axis of the body frame and $N^{i}$, and $\phi_{N}$ is the angle that the projection of $N^{i}$ makes in the XY-plane with the X-axis. We consider these angles to be constant in the observer's frame. The projection into the TT gauge can be performed via Eq.~(11.44) in~\cite{PoissonWill}, which gives us the following plus and cross polarizations for the waveform,
\begin{align}
    \label{eq:h-time}
    h &= h_{+} - i h_{\times} 
    \nonumber \\
    &= \frac{\nu M}{D_{L}} \tilde{u}^{2} \sum_{mn} A_{m,n}(\iota,\Omega) e^{in\phi} {_{-2}}Y_{2m}(\theta_{N},\phi_{N})
\end{align}
where ${_{-2}}Y_{lm}(\theta,\phi)$ are spin weight $-2$ spherical harmonics, $m$ is an integer such that $|m|\le2$ and $n=\pm2$. The amplitude functions $A_{m,n}$ are listed in Appendix~\ref{app:wavamps}.

Since the binary is precessing, the amplitudes $A_{mn}$ depend on time through $[\iota,\Omega]$. In order to calculate the Fourier domain waveform, we make use of the SPA and SUA~\cite{Klein:2014bua} to obtain the precession corrections. The phase of the Fourier integral is of the standard form $\Psi_{F} = 2\pi f t(\tilde{u}) + n \phi(\tilde{u})$, and the stationary point is given by $\tilde{u}_{\star} = \tilde{u}(t_{\star}) = (-2\pi M f/n)^{1/3}$. Note that this only contributes to the Fourier transform for positive frequencies for $n=-2$. The SUA corrections are found through $T_{n} = 1/\sqrt{n \ddot{\phi}(\tilde{u}_{\star})}$. After applying both the SPA and SUA corrections, the resulting waveform is
\begin{align}
    \label{eq:h-freq}
    \tilde{h}(f) &= \sqrt{\frac{5}{96}} \frac{{\cal{M}}^{5/6}}{\pi^{2/3}D_{L}} f^{-7/6} e^{i\tilde{\Psi}_{F}}  \sum_{m} {\cal{A}}_{m}(f) {_{-2}}Y_{2m}(\theta_{N}, \phi_{N})
\end{align}
where the Fourier phase is
\begin{align}
    \label{eq:wave_phase}
    \tilde{\Psi}_{F} &= 2\pi f t_{c} - 2\phi_{c} - \frac{\pi}{4} 
    \nonumber \\
    &+ \frac{3}{128 \nu (\pi M f)^{5/3}} \left[1 - \frac{5Q_{0}}{4M^{3} \nu }\sqrt{\frac{\pi}{5}} \epsilon_{1}^{p} \epsilon_{2}^{q} \; {\cal{U}}_{pq} \left(\pi M f\right)^{4/3} \right]\,,
\end{align}
and the Fourier amplitudes are
\begin{align}
    \label{eq:wave_amps}
    {\cal{A}}_{m}(f) = \sum_{k=0}^{k_{\rm max}} \frac{a_{k,k_{\rm max}}}{2}\left\{ A_{m,-2}\left[\psi_{2}(\tilde{u}_{k})\right] + A_{m,-2}\left[\psi_{2}(\tilde{u}_{-k})\right]\right\}\,,
\end{align}
which are dependent on frequency through the SUA corrected $\tilde{u}_{k}$, specifically
\begin{align}
    \tilde{u}_{k} &= \tilde{u}(t_{\star} + k T_{n})
    \nonumber \\
    &= (\pi M f)^{1/3} + 4 k \sqrt{\frac{\nu}{15}} (\pi M f)^{7/6}
    \nonumber \\
    &\times \left[1 + \frac{Q_{0}}{16M^{3} \nu }\sqrt{\frac{\pi}{5}} \epsilon_{1}^{p} \epsilon_{2}^{q} \; {\cal{U}}_{pq} (\pi M f)^{4/3}\right]\,.
\end{align}
In the above, the coefficients $a_{k,k_{\rm max}}$ satisfy the linear system of equations
\begin{equation}
    \frac{(-i)^{p}}{2^{p} p!} = \sum_{k=0}^{k_{\rm max}} a_{k,k_{\rm max}} \frac{k^{2p}}{(2p)!}\,,
\end{equation}
for $p\in{0,...,k_{\rm max}}$. The value of $k_{\rm max}$ is usually chosen based on the desired level of faithfulness when compared to numerical waveforms, as well as computational efficiency~\cite{Klein:2014bua}. This completes the Fourier domain waveform for generic mass quadrupole effects.

\begin{figure}[htb!]
    \centering
    \includegraphics[width=\columnwidth, trim={2cm, 8cm, 2cm, 8cm}, clip]{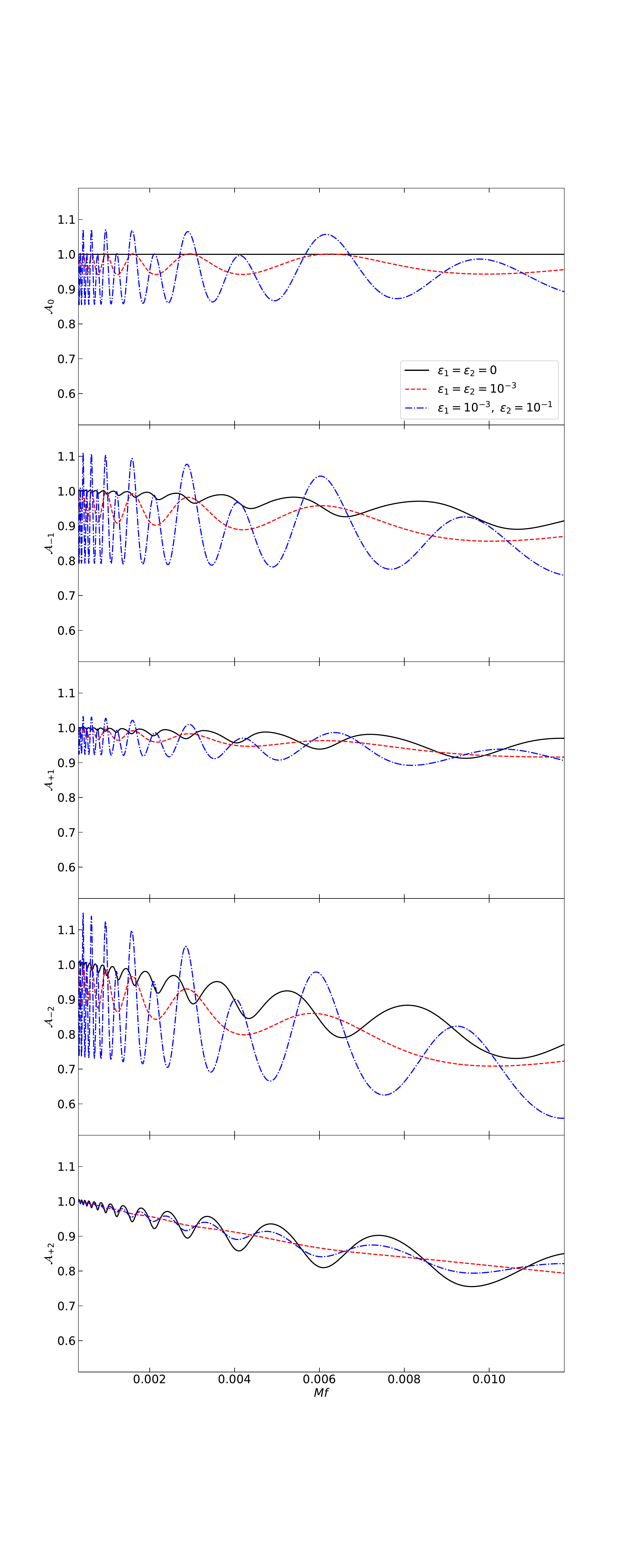}
    \caption{Comparison of the waveform amplitudes ${\cal{A}}_{m}$ from Eq.~\eqref{eq:wave_amps} for the spheroidal case (black), $\epsilon_{1} = \epsilon_{2} = 10^{-3}$ (red, dashed), and $\epsilon_{1} = 10^{-3}, \epsilon_{2} = 10^{-1}$ (blue, dot-dashed). The amplitudes are all normalized such that ${\cal{A}}_{m} = 1$ at the lowest frequency plotted.}
    \label{fig:amps}
\end{figure}

To showcase these waveforms, we plot the amplitude functions ${\cal{A}}_{m}(f)$ in Fig.~\ref{fig:amps} for different values of the modulus parameters $\epsilon_{m}$, specifically $\epsilon_{m} = 0$ (black lines) which corresponds to the spheroidal configuration, $\epsilon_{m} = 10^{-3}$ (red dashed lines), and $\epsilon_{2} = 10^{-1} = 100 \epsilon_{1}$. The amplitude functions are normalized such that ${\cal{A}}_{m}(f_{\rm low}) = 1$, where $(\pi M f_{\rm low})^{1/3} = 0.1$.  For the spheroidal case, the inclination angle $\iota$ becomes a constant, and thus ${\cal{A}}_{0}$ also becomes constant with frequency. The amplitudes functions are generally modulated due to the precession of the orbital angular momentum, which defines the axis along which the GW amplitude is largest.

\begin{figure}[htb!]
    \centering
    \includegraphics[width=\columnwidth, trim=1.5cm 2cm 2cm 2cm, clip]{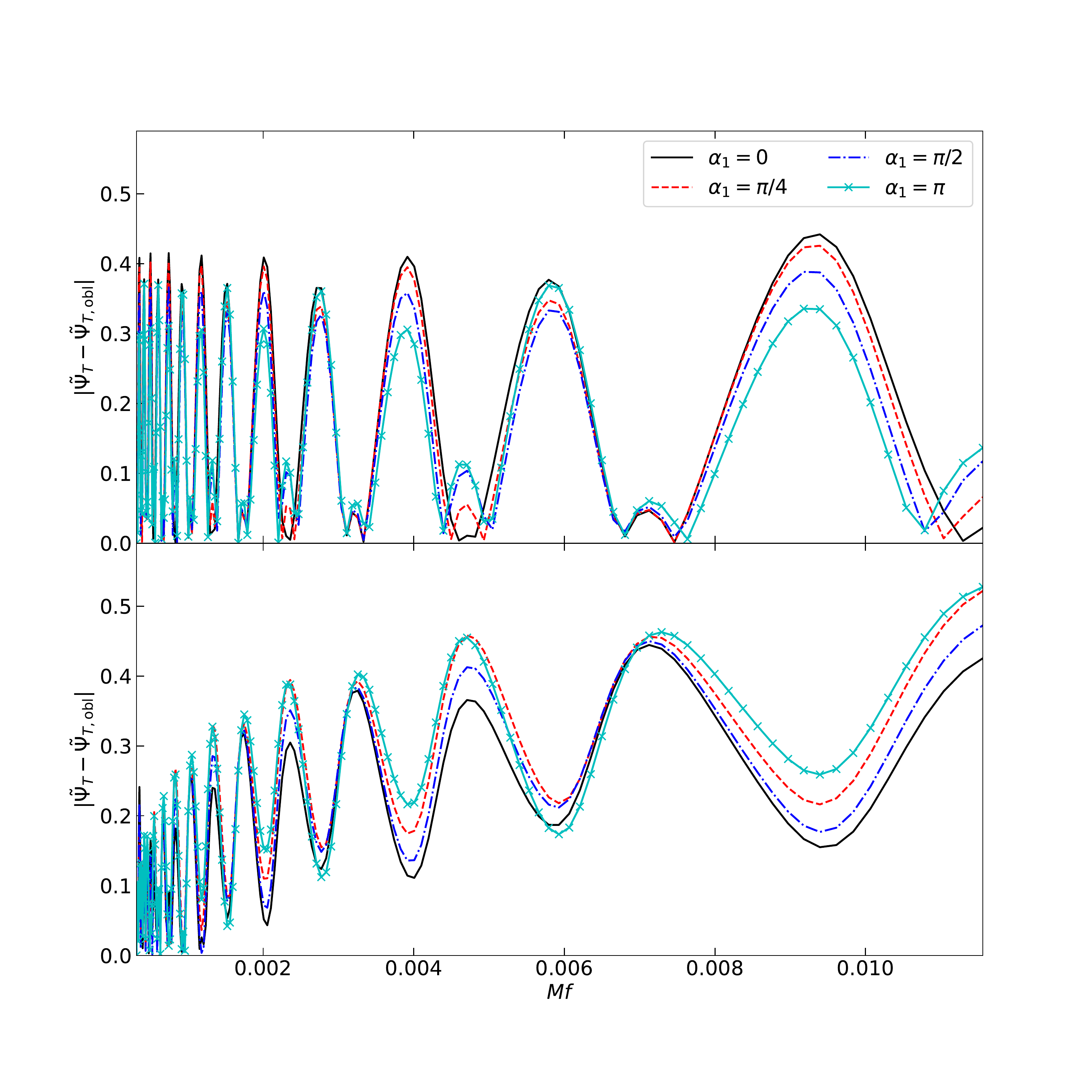}
    \caption{Top Panel: Total GW phase difference between an oblate/prolate configuration and one with $\epsilon_{1} = \epsilon_{2} = 10^{-3}$. The polar phase $\alpha_{2} = 0$, while the different lines correspond to different values of $\alpha_{1}$, namely 0 (black), $\pi/4$ (red, dashed), $\pi/2$ (blue, dot-dashed), and $\pi$ (cyan, crossed). Bottom Panel: The same as the top panel but with $\epsilon_{1} = 10^{-3}$ and $\epsilon_{2} = 10^{-1}$.}
    \label{fig:phase}
\end{figure}

Finally, in Fig.~\ref{fig:phase} we plot the total phase difference between a spheroidal configuration and two configurations with $\epsilon_{m} = 10^{-3}$ (top panel) and $\epsilon_{2} = 10^{-1} = 100 \epsilon_{1}$ (bottom panel). The total phase of the waveform is found by
\begin{equation} \label{phiTfinal}
    \tilde{\Psi}_{T}(f) = \tilde{\Psi}_{F}(f) + {\rm arg}\left[\sum_{m}{\cal{A}}_{m}(f)\right]\,.
\end{equation}
The different lines in each panel correspond to different values of $\alpha_{1}$, while $\alpha_{2} = 0$ for all cases. 

As a simplistic but useful rule of thumb, an effect introducing a phase difference of $0.1$ or greater is likely to substantially impact a matched-filter search, leading to a significant loss of detected events if the matched-filter search is done with waveforms that do not include these corrections~\cite{Lindblom:2008cm}. Or, in other words, generally a phase difference of $0.1$ would in principle be observable by the LIGO and Virgo detectors at signal-to-noise ratio 10. Therefore, for all the cases shown in Fig.~\ref{fig:phase}, the deviations from spheroidness could be detectable, although we stress that this naive estimates must be validated with a detailed parameter estimation study, also taking into account possible parameter correlations and systematic errors.

\section{Discussion and outlook}

We have here developed the first analytic waveforms to model general mass quadrupole moment effects of compact objects. The waveforms are parameterized by the quadrupole parameter $Q_{0}$ corresponding to the oblate/prolate configuration, and the modulus $\epsilon_{m}$ and phase $\alpha_{m}$ parameters that describe deviations from spheroidness. The latter of these are generic enough for us to consider constraining  non-axisymmetric configurations of compact objects. 
Besides considering non-axisymmetric bodies, the tools developed here can also be used to compute the leading-order correction for current quadrupole moments. This was partially addressed in Ref.~\cite{Fransen:2022jtw} but only for the axisymmetric case and the main results were obtained in the EMRI limit.  
It should be stressed though that the generic mass quadrupole corrections considered here also break the equatorial symmetry and affect the waveform at a lower PN order relative to the axisymmetric ${\cal S}_2$ corrections considered in~\cite{Fransen:2022jtw}. They should therefore be the leading-order signatures for generic objects without equatorial symmetry.
Another natural extension of our work is to include the effect of the objects' angular momenta, which can give rise to a variety of phenomena (e.g., spin precession and coupling to the quadrupole moment).

A crucial aspect that we did not address in this paper is the extent with which GW detectors will be able to constrain or detect non-axisymmetric mass quadrupole moments. Based on a dephasing argument, we estimate that even small deviations from spheroidness might be measurable with current generation ground-based detectors. However, such an argument does not take into account the correlations among the physical parameters of the binary, or the possibility of degeneracies that would limit our ability to stringently constrain the additional quadrupole parameters. 
One degeneracy that can already be seen in the analysis carried out here is the fact that the waveforms depend on the components of the effective quadrupole moment tensor defined below Eq.~\eqref{eq:LMQ}, and not on the individual quadrupole moments of the objects. This is not surprising given that a similar situation happens when considering the leading PN spin and tidal corrections to the waveform~\cite{Flanagan:2007ix}. 
In our specific case the situation is even worse, since the quadrupole tensor enters at the leading PN order through the \emph{combination} $\epsilon_{1}^{p} \epsilon_{2}^{q} {\cal{U}}_{pq}$ in Eq.~\eqref{eq:wave_phase}, so individual quadrupole components are degenerated.
For example, based solely on the 2PN inspiral corrections, a (admittedly fine-tuned) model in which different components of the quadrupole moments conspire to give a negligible deviation from the standard Kerr case cannot be excluded. Higher-order PN corrections in the phase and amplitude can break this degeneracy.

A rough estimate of the constraints on generic quadrupolar deformations can come from measured upper bounds on the parametrized corrections, $\delta \phi_2$, to the 2PN coefficients. For the neutron-star binary GW170817 such constraints read $|\delta \phi_2|\lesssim 3.5$ at $90\%$
confidence level~\cite{LIGOScientific:2018dkp}. For BH binaries, the latest bound obtained by combining all GWTC-3 events reads $|\delta \phi_2|\lesssim 0.1$~\cite{LIGOScientific:2021sio} (assuming the same type of deviations for all sources). These measurements could roughly translate into an upper bound on the combination $\epsilon_{1}^{p} \epsilon_{2}^{q} {\cal{U}}_{pq}$ (of both binary components) in Eq.~\eqref{eq:wave_phase}. However, such bounds were derived without taking into account the amplitude corrections (see Eq.~\eqref{eq:wave_amps}) and the amplitude modulation (see Eq.~\eqref{phiTfinal}) found in this work, so a detailed analysis should be performed to obtain faithful constraints.
At any rate, the order of magnitude of these constraints makes such parameter estimation a promising future avenue.

Another open question is how do the new parameters, specifically $[\epsilon_{m}, \alpha_{m}]$, map to the properties of compact objects. For BHs, $\epsilon_{m}=0$, but in general this need not be true. In general, these parameters will be equation of state dependent. Having specific theoretical predictions of these values for various astrophysical and exotic compact objects would allow one to map from generic constraints on the modulus and phase parameters to the physical set of parameters that characterizes the equation of state and structure of the bodies.

The calculations that we have carried out here are the first step toward more general investigations of the structure of compact objects, and we plan to investigate the above topics in future work.

\begin{acknowledgments}
We acknowledge financial support provided under the European Union's
H2020 ERC, Starting Grant agreement no.~DarkGRA--757480.
Computations were performed at Sapienza University of Rome on the Vera
cluster of the Amaldi Research Center.
This project has received funding from the European Union's Horizon
2020 research and innovation programme under the Marie
Sk\l odowska-Curie grant agreement No 101007855.
We also acknowledge support under the MIUR PRIN and FARE programmes (GW-NEXT,
CUP:~B84I20000100001,  2020KR4KN2), and from the Amaldi Research
Center funded by the MIUR program ``Dipartimento di Eccellenza''
(CUP:~B81I18001170001). This work is partially supported by the PRIN
Grant 2020KR4KN2 ``String Theory as a bridge between Gauge Theories
and Quantum Gravity''.
R.B. acknowledges financial support provided by FCT -- Funda\c{c}\~{a}o para a Ci\^{e}ncia e a Tecnologia, I.P., under the Scientific Employment Stimulus -- Individual Call -- 2020.00470.CEECIND.
The authors would like to acknowledge networking support by the GWverse COST Action
CA16104, ``Black holes, gravitational waves and fundamental physics.''
\end{acknowledgments}
\ \\

\appendix
\section{Higher PN Effects}
\label{pn}

To complete the discussion in Sec.~\ref{sec:notation}, we here provide some relevant PN quantities. We do not formally use these in the analysis presented in this paper. The 1PN and 2PN order corrections to the point particle Lagrangian in Eq.~\eqref{eq:LM} are
\begin{align}
    {\cal{L}}_{\rm 1PN} &= \frac{1-3 \nu}{8}v^4 + \frac{M}{2r} \left[\left(3+ \nu\right)v^2+ \nu
  \dot{r}^2-\frac{M}{r} \right]\,,
  \\
  {\cal{L}}_{\rm 2PN} &= \frac{1}{16}(1-7\nu+13\nu^{2}) v^{6} + \frac{M}{8r}\left[(7-12\nu-9\nu^{2})v^{4} 
  \right.
  \nonumber\\
  &\left.+ (4-10\nu)\nu\dot{r}^{2}v^{2} + 3\nu^{2}\dot{r}^{4}\right] + \frac{M^{2}}{2r^{2}}\left[(4-2\nu+\nu^{2})v^{2}
  \right.
  \nonumber\\
  &\left.+3\nu(1+\nu)\dot{r}^{2}\right] + \frac{M^{3}}{4r^{3}}\,,
\end{align}
The PN corrections to the conserved orbital energy are
\allowdisplaybreaks[4]
\begin{align}
    \frac{E_{\rm 1PN}}{\mu} &= \frac{3}{8} (1-3\nu) v^{4} + \frac{1}{2} (3+\nu) v^{2} \frac{M}{r} + \frac{1}{2} \nu \frac{M}{r} \dot{r}^{2} 
    \nonumber \\
    &+ \frac{1}{2} \left(\frac{M}{r}\right)^{2}\,,
    \\
    \frac{E_{\rm 2PN}}{\mu} &= \frac{5}{16} (1-7\nu + 13\nu^{2}) v^{6} + \frac{1}{8}(21-23\nu-27\nu^{2})\frac{M}{r} v^{4}
    \nonumber \\
    &+ \frac{1}{4} \nu (1-15\nu) \frac{M}{r} v^{2} \dot{r}^{2} - \frac{3}{8} \nu(1-3\nu)\frac{M}{r} \dot{r}^{4} 
    \nonumber \\
    &+ \frac{1}{8}(14-55\nu+4\nu^{2})\left(\frac{M}{r}\right)^{2}v^{2}  
    \nonumber \\
    &+ \frac{1}{8}(4+69\nu+12\nu^{2})\left(\frac{M}{r}\right)^{2}\dot{r}^{2} - \frac{1}{4}(2+15\nu) \left(\frac{M}{r}\right)^{3}\,,
    \\
    E_{\rm SO} &= \frac{\mu}{r^{2}} \epsilon^{ijk} \Sigma^{i} n^{j} v^{k}\,,
    \\
    E_{\rm SS} &= \frac{1}{r^{3}} \left[3 (n^{i} S_{1}^{i}) (n^{j} S_{2}^{j}) - S_{1}^{i} S_{2}^{i}\right]\,.
\end{align}
Finally, the corrections to the conserved orbital angular momentum are
\allowdisplaybreaks[4]
\begin{align}
    L_{\rm 1PN} &= \frac{1}{2} v^{2} \left(1-3\nu\right) + \left(3+\nu\right) \frac{M}{r}\,, 
    \\
    L_{\rm 2PN} &= (1-7\nu+13\nu^{2})\frac{3}{8} v^{4} + (7-10\nu-9\nu^{2}) \frac{M}{2r}v^{2} 
    \nonumber \\
    &- \frac{1}{2} \nu (2+5\nu) \frac{M}{r}\dot{r}^{2} + \frac{1}{4} (14-41\nu+4\nu^{2}) \left(\frac{M}{r}\right)^{2}\,,
    \\
    L^{i}_{\rm SO} &= \frac{\mu}{M} \left[\frac{M}{r} \epsilon^{ijk} \epsilon^{kpq} n^{j} n^{p} (2S^{q} + \Sigma^{q}) 
    \right.
    \nonumber \\
    &\left.
    - \frac{1}{2} \epsilon^{ijk} \epsilon^{kpq} v^{j} v^{p} \Sigma^{q}\right]\,.
\end{align}

\section{PN Quadrupole Coefficients}
\label{app:coeffs}

We here provide the coefficients of leading PN order corrections to various orbital quantities derived in Sec.~\ref{sec:oscorb} and radiation reaction effects in Sec.~\ref{sec:rr} due to generic mass quadrupole moments. With $Q^{a} = [Q_{0}, Q^{R}_{1}, Q^{I}_{1}, Q^{R}_{2}, Q^{I}_{2}]$, the coefficients in Eqs.~\eqref{eq:e-new}-\eqref{eq:p-new} are
\allowdisplaybreaks[4]
\begin{widetext}
\begin{align}
    \tilde{\cal{E}}_{0} &= 6 + 2 e_{0}^2 + 
 6 (3 + e_{0}^2) \cos(2 \iota) - (5 + 6 e_{0}
 + e_{0}^2) \cos[2 (\iota - \omega)] + 10 \cos(2 \Omega) +
 12 e_{0}\cos(2 \omega) + 2 e_{0}^2 \cos(2 \omega) 
 \nonumber \\
 &- 5 \cos[2 (\iota + \omega)] - 6 e_{0} \cos[2 (\iota + \omega)] - 
 e_{0}^2 \cos[2 (\iota + \omega)]\,,
 \\
 \tilde{\cal{E}}_{1} &= 8 \sqrt{\frac{2}{3}} \sin\iota \left\{(5 + 6 e_{0} + e_{0}^{2}) \cos\Omega \sin(2\omega) + \cos\iota \left[-3(3+e_{0}^{2}) + (5+6e_{0}+e_{0}^{2})\cos(2\omega)\right]\sin\Omega \right\}\,,
 \\
 \tilde{\cal{E}}_{2} &= 2 \sqrt{\frac{2}{3}} \sin\iota \left\{4 \cos\iota \left[-3(3+e_{0}^{2}) + (5 + 6e_{0} + e_{0}^{2}) \cos(2\omega) \right]\cos\Omega - 4 (5+6e_{0}+e_{0}^{2}) \sin(2\omega)\sin\Omega  \right\}\,,
 \\
 {\tilde{\cal{E}}}_{3} &= \frac{1}{\sqrt{6}} \left( -2 \cos(2\Omega) \left\{2 (1+e_{0}) (5+e_{0}) [3 + \cos(2\iota)] \cos(2\omega) + 12 (3+e_{0}^{2}) \sin^{2}\iota\right\} + 16 (1+e_{0}) (5+e_{0}) \cos\iota \sin(2\omega) \sin(2\Omega)\right)\,,
 \\
 {\tilde{\cal{E}}}_{4} &= 2 \sqrt{\frac{2}{3}} \left\{4(1+e_{0})(5+e_{0}) \cos\iota \cos(2\Omega) \sin(2\omega) + \left[(1+e_{0})(5+e_{0})(3 + \cos(2\iota)) \cos(2\omega) + 6 (3+e_{0}^{2})\sin^{2}\iota\right]\sin(2\Omega)\right\}\,,
 \\
 {\tilde{\cal{P}}}_{0} &= (3+e_{0})^{2} \left[1 + 3 \cos(2\iota)\right] + 2 (1+e_{0}) (5+e_{0}) \cos(2\omega) \sin^{2}\iota\,,
 \\
 {\tilde{\cal{P}}}_{1} &= 4 \sqrt{\frac{2}{3}} \sin\iota \left\{(1+e_{0})(5+e_{0})\cos\Omega \sin(2\omega) + \cos\iota \left[-3(3+e_{0}^{2}) + (1+e_{0})(5+e_{0}) \cos(2\omega)\right]\sin\Omega\right\}\,,
 \\
 \tilde{\cal{P}}_{2} &= \sqrt{\frac{2}{3}} \sin\iota \left\{ 4 \cos\iota \left[-3(3+e_{0}^{2}) + (1+e_{0})(5+e_{0}) \cos(2\omega)\right]\cos\Omega - 4(1+e_{0})(5+e_{0}) \sin(2\omega) \sin\Omega\right\}\,,
 \\
 \tilde{\cal{P}}_{3} &= \frac{1}{2\sqrt{6}} \left\{-2 \cos(2\Omega) \left[2(1+e_{0})(5+e_{0}) (3 + \cos(2\iota))\cos(2\omega) + 12 (3+e_{0}^{2}) \sin^{2}\iota\right] +16(1+e_{0})(5+e_{0}) \cos\iota \sin(2\omega) \sin(2\omega)\right\}\,,
 \\
 \tilde{\cal{P}}_{4} &= \sqrt{\frac{2}{3}} \left\{4 (1+e_{0})(5+e_{0}) \cos\iota \cos(2\Omega) \sin(2\omega) + \left[(1+e_{0})(5+e_{0}) (3 + \cos(2\iota)) \cos(2\omega) + 6(3+e_{0}^{2}) \sin^{2}\iota\right] \sin(2\Omega)\right\}\,.
\end{align}
\end{widetext}
The coefficients of the modified Kepler's third law in Eq.~\eqref{eq:kep3} are
\allowdisplaybreaks[4]
\begin{align}
    \tilde{\Omega}_{0} &= 12 \left[3 + 9 \cos(2\iota) - \cos(2\omega) \sin^{2}\iota\right]\,,
    \\
    \tilde{\Omega}_{1} &= -4\sqrt{6} \left\{2 \cos\Omega \sin\iota \sin(2\omega) \right.
    \nonumber \\
    &\left.
    + \left[18 + \cos(2\omega)\right] \sin(2\iota) \sin\Omega\right\}\,,
    \\
    \tilde{\Omega}_{2} &= 4 \sqrt{6} \left\{36 \cos\iota \cos\Omega \sin\iota - \left[36 + \cos(2\omega)\right] \cos\Omega \sin(2\iota) \right.
    \nonumber \\
    &\left.
    + 2 \sin\iota \sin(2\omega) \sin\Omega\right\}
    \\
    \tilde{\Omega}_{3} &= \sqrt{6} \left\{ \cos(2\Omega) \left[2 \left(3 + \cos(2\iota)\right] \cos(2\omega) - 72 \sin^{2}\iota\right] \right.
    \nonumber \\
    &\left.
    - 8 \cos\iota \sin(2\omega) \sin(2\Omega)\right\}\,,
    \\
    \tilde{\Omega}_{4} &= 2 \sqrt{6} \left\{-4 \cos\iota \cos(2\Omega) \sin(2\omega) \right.
    \nonumber \\
    &\left.
    - \left[\left(3+\cos(2\iota)\right)\cos(2\omega) - 36 \sin^{2}\iota\right]\sin(2\Omega)\right\}\,.
\end{align}
Lastly, the coefficients for the orbital energy in Eq.~\eqref{eq:E-os} are
\allowdisplaybreaks[4]
\begin{align}
    \tilde{E}_{0} &= 18 \cos(2\omega) \sin^{2}\iota\,,
    \\
    \tilde{E}_{1} &= 6 \sqrt{6} \left[ \cos(2\omega) \sin(2\iota) + 2 \cot\Omega \sin\iota \sin(2\omega)\right] \sin\Omega\,,
    \\
    \tilde{E}_{2} &= 12 \sqrt{6} \sin\iota \left[\cos\iota \cos(2\omega) \cos\Omega - 2 \cos\omega \sin \omega \sin\Omega\right]\,,
    \\
    \tilde{E}_{3} &= 3 \sqrt{6} \left\{- \left[\left(3 + \cos(2\iota)\right) \cos(2\omega) \cos(2\Omega)\right] 
    \right.
    \nonumber \\
    &\left.
    + 4 \cos\iota \sin(2\omega) \sin(2\Omega)\right\}\,,
    \\
    \tilde{E}_{4} &= 3 \sqrt{6} \left\{4 \cos\iota \cos(2\Omega) \sin(2\omega) 
    \right.
    \nonumber \\
    &\left.
    + \left[3 + \cos(2\iota)\right] \cos(2\omega) \sin(2\Omega)\right\}\,.
\end{align}
The corrections to the evolution of the orbital frequency due to radiation reaction in Eq.~\eqref{eq:dudt} are
\allowdisplaybreaks[4]
\begin{align}
    \tilde{U}_{0} &= -13 - 39 \cos(2\iota) + 98 \cos(2\omega) \sin^{2}\iota\,,
    \\
    \tilde{U}_{1} &= 2\sqrt{\frac{2}{3}} \left\{98 \cos\Omega \sin\iota \sin(2\omega) 
    \right.
    \nonumber \\
    &\left.
    + \left[39 + 49 \cos(2\omega)\right] \sin(2\iota) \sin\Omega\right\}\,,
    \\
    \tilde{U}_{2} &= 2 \sqrt{\frac{2}{3}} \left\{\left[39+49 \cos(2\omega)\right] \cos\Omega \sin(2\iota) 
    \right.
    \nonumber \\
    &\left.
    - 98 \sin\iota \sin(2\omega) \sin\Omega \right\}\,,
    \\
    \tilde{U}_{3} &= \sqrt{\frac{2}{3}} \left\{\cos(2\Omega) \left[-49 \left(3 + \cos(2\iota)\right) \cos(2\omega) + 78 \sin^{2}\iota\right] 
    \right.
    \nonumber \\
    &\left.
    + 196 \cos\iota \sin(2\omega) \sin(2\Omega)\right\}\,,
    \\
    \tilde{U}_{4} &= \sqrt{\frac{2}{3}} \left\{196 \cos\iota \cos(2\Omega) \sin(2\omega) 
    \right.
    \nonumber \\
    &\left.
    + \left[49 \left(3 + \cos(2\iota)\right] \cos(2\omega) - 78 \sin^{2}\iota\right]\sin(2\Omega)\right\}\,.
\end{align}

To obtain the coefficients ${\cal{U}}_{pq}$ in Eq.~\eqref{eq:avg-coeffs}, one has to compute the averages of the $\tilde{U}_{a}$ coefficients listed above in a small $\epsilon_{1,2}$ expansion. The $\tilde{U}_{a}$ are coupled to the quadrupole coefficients $Q_{a}$, and so each $\tilde{U}_{a}$ must be expanded to different orders. For brevity, we only list the ${\cal{U}}_{pq}$ up to linear order in the expansion. This means that $\tilde{U}_{0}$ must be computed to ${\cal{O}}(\epsilon_{1}, \epsilon_{2})$ with remainders of order ${\cal{O}}(\epsilon_{1}^{2}, \epsilon_{2}^{2}, \epsilon_{1} \epsilon_{2})$, while all other $\tilde{U}_{a}$ must be computed to ${\cal{O}}(\epsilon_{1}^{0}, \epsilon_{2}^{0})$. The reason for this is that these are coupled to $Q^{R,I}_{m}$ which are already linear in $\epsilon_{m}$, i.e. $Q^{R,I}_{m} \sim \epsilon_{m} Q_{0}$. Writing $\tilde{U}_{0} = \tilde{U}_{0}^{(0)} + \epsilon_{1} \tilde{U}_{0}^{(1)} + \epsilon_{2} \tilde{U}_{0}^{(2)} + {\cal{O}}(\epsilon_{1}^{2}, \epsilon_{2}^{2}, \epsilon_{1} \epsilon_{2})$, the end results are
\begin{align}
    {\cal{U}}_{00} &= -13 -39 \cos(2\iota_{0}) 
    \nonumber \\
    &+ \frac{196}{\pi \zeta} \cos\left(\frac{\pi\zeta}{2} -2 \omega_{0}\right) \sin\left(\frac{\pi\zeta}{2}\right) \sin^{2}\iota_{0}
    \\
    {\cal{U}}_{10} &= \langle \tilde{U}_{0}^{(1)} \rangle_{\psi_{2}} + \sqrt{\frac{3}{2}} \left[\cos\alpha_{1} \langle \tilde{U}_{1} \rangle_{\psi_{2}} + \sin\alpha_{1} \langle \tilde{U}_{2} \rangle _{\psi_{2}} \right]
    \\
    {\cal{U}}_{01} &= \langle \tilde{U}_{0}^{(2)} \rangle_{\psi_{2}} + \sqrt{\frac{3}{2}} \left[\cos(2\alpha_{2}) \langle \tilde{U}_{3} \rangle_{\psi_{2}} + \sin(2\alpha_{2}) \langle \tilde{U}_{4} \rangle _{\psi_{2}} \right]
\end{align}
where $\zeta = [3 + 5 \cos(2\iota_{0})]\sec\iota_{0}$, and
\allowdisplaybreaks[4]
\begin{widetext}
\begin{align}
    \langle \tilde{U}_{0}^{(1)} \rangle_{\psi_{2}} &= \frac{1}{2\pi\zeta^{2}} \Bigg(\frac{49\zeta}{\zeta^{2}-4}\sin\left[\frac{\pi\zeta}{2}\right] \left\{8 \sin(2\iota_{0})\left[\left(\zeta+2\right)\sin\left(\frac{\pi\zeta}{2} -\Delta -2 \omega_{0}\right) + \left(\zeta-2\right) \sin\left(\frac{\pi\zeta}{2}+\Delta - 2\omega_{0}\right)\right] 
    \right.
    \nonumber \\
    &\left.
    + \left[7 + 4 \cos(2\iota_{0}) + 5 \cos(4\iota_{0})\right] \sec\iota_{0} \tan\iota_{0} \left[\left(\zeta+2\right)\sin\left(\frac{\pi\zeta}{2} -\Delta -2 \omega_{0}\right) - \left(\zeta-2\right) \sin\left(\frac{\pi\zeta}{2}+\Delta - 2\omega_{0}\right)\right]\right\} 
    \nonumber \\
    &+ 4\sin\Delta \left[39 \pi \zeta \sin\{2\iota_{0}\} - 49 \{7 + 5 \cos(2\iota_{0})\} \sin\iota_{0} \{\pi \zeta \cos(\pi \zeta - 2\omega_{0}) - \sin(\pi \zeta - 2\omega_{0}) - \sin(2\omega_{0})\} \tan^{2}\iota_{0} \right]\Bigg)\,,
    \\
    \langle \tilde{U}_{0}^{(2)} \rangle_{\psi_{2}} &= \frac{1}{8\pi \zeta(\zeta-4)(\zeta+4)} \left\{16 \zeta \sin^{2}\iota_{0} \left[-39\pi \zeta (\zeta^{2}-16) + 784 \sin(\pi\zeta-2\omega_{0}) + 784 \sin(2\omega_{0})\right] 
    \right.
    \nonumber \\
    &\left.
    + 98 \sec\iota_{0} \left[\pi \zeta (\zeta^{2}-16)(23 + 36\cos(2\iota_{0}) + 5\cos(4\iota_{0})\right]\cos(\pi\zeta-2\omega_{0}) 
    \right.
    \nonumber \\
    &\left.
    - 2\left[-184+3\zeta^{2}+8(\zeta^{2}-36)\cos(2\iota_{0})+5(\zeta^{2}-8)\cos(4\iota_{0})\right]\left[\sin(\pi\zeta-2\omega_{0})+\sin(2\omega_{0})\right]\tan^{2}\iota_{0}\right\}\,,
    \\
    \langle \tilde{U}_{1}\rangle_{\psi_{2}} &= -\frac{98}{\pi(\zeta^{2}-4)}\sqrt{\frac{2}{3}} \sin\left(\frac{\pi\zeta}{2}\right)\left\{(\zeta+2)\left[2\sin\iota_{0} + \sin(2\iota_{0})\right] \sin\left(\alpha_{2}-2\omega_{0} + \frac{\pi\zeta}{2}\right) 
    \right.
    \nonumber \\
    &\left.
    - (\zeta-2) \left[2\sin\iota_{0} - \sin(2\iota_{0})\right]\sin\left(\alpha_{2} + 2\omega_{0} - \frac{\pi\zeta}{2}\right)\right\}
    \\
    \langle \tilde{U}_{2} \rangle_{\psi_{2}} &= \frac{98}{\pi(\zeta^{2}-4)} \sqrt{\frac{2}{3}} \sin\left(\frac{\pi\zeta}{2}\right) \left\{(\zeta+2) \left[2\sin\iota_{0} + \sin(2\iota_{0})\right] \cos\left(\alpha_{2} - 2\omega_{0} + \frac{\pi\zeta}{2}\right) 
    \right.
    \nonumber \\
    &\left.
    - (\zeta-2) \left[2\sin\iota_{0} - \sin(2\iota_{0})\right] \cos\left(\alpha_{2} + 2\omega_{0} - \frac{\pi\zeta}{2}\right)\right\}\,, 
    \\
    \langle \tilde{U}_{3}\rangle_{\psi_{2}} &= - \frac{392}{\pi (\zeta^{2}-16)} \sqrt{\frac{2}{3}} \sin\left(\frac{\pi\zeta}{2}\right) \left[(\zeta+4) \cos^{4}\left(\frac{\iota_{0}}{2}\right)\cos\left(2\alpha_{2} - 2\omega_{0} + \frac{\pi\zeta}{2}\right) 
    \right.
    \nonumber \\
    &\left.
    + (\zeta-4)\sin^{4}\left(\frac{\iota_{0}}{2}\right)\cos\left(2\alpha_{2} + 2\omega_{0} - \frac{\pi\zeta}{2}\right)\right]\,,
    \\
    \langle \tilde{U}_{4} \rangle_{\psi_{2}} &= - \frac{392}{\pi (\zeta^{2}-16)} \sqrt{\frac{2}{3}} \sin\left(\frac{\pi\zeta}{2}\right) \left[(\zeta+4) \cos^{4}\left(\frac{\iota_{0}}{2}\right)\sin\left(2\alpha_{2} - 2\omega_{0} + \frac{\pi\zeta}{2}\right) 
    \right.
    \nonumber \\
    &\left.
    + (\zeta-4)\sin^{4}\left(\frac{\iota_{0}}{2}\right)\sin\left(2\alpha_{2} + 2\omega_{0} - \frac{\pi\zeta}{2}\right)\right]\,.
\end{align}
\end{widetext}

\section{Waveform Amplitudes}
\label{app:wavamps}

We here provide the waveform amplitudes $A_{m,n}^{+,\times}(\iota,\Omega)$ from Eq.~\eqref{eq:h-time}.
\begin{align}
    A_{0,\pm2} & = 2\sqrt{\frac{6\pi}{5}} \sin^{2}\iota\,,
    \\
    A_{+1,+2} &= - (A_{-1,-2})^{\dagger} =  8i\sqrt{\frac{\pi}{5}} e^{-i\Omega} \sin\iota \sin^{2}(\iota/2)\,,
    \\
    A_{+1,-2} &= - (A_{-1,+2})^{\dagger} = -16i\sqrt{\frac{\pi}{5}} e^{-i\Omega} \sin(\iota/2) \cos^{3}(\iota/2)\,,
    \\
    A_{+2,+2} &= (A_{-2,-2})^{\dagger} = -8\sqrt{\frac{\pi}{5}} e^{-2i\Omega} \sin^{4}(\iota/2)\,,
    \\
    A_{+2,-2}&= (A_{-2,+2})^{\dagger} = -8\sqrt{\frac{\pi}{5}} e^{-2i\Omega} \cos^{4}(\iota/2)\,.
\end{align}
%

\bibliography{refs}
\end{document}